\documentclass[twocolumn]{aastex63}

\newcommand\teff{\mbox{$T_\mathrm{eff}$}}

\newcommand\mjup{\mbox{$M_\mathrm{Jup}$}}

\usepackage{color,hyperref}

\begin{document}

\title{Eight New Substellar Hyades Candidates from the UKIRT Hemisphere Survey}

\correspondingauthor{Adam C. Schneider}
\email{adam.c.schneider4.civ@us.navy.mil}

\author[0000-0002-6294-5937]{Adam C. Schneider}
\affil{United States Naval Observatory, Flagstaff Station, 10391 West Naval Observatory Rd., Flagstaff, AZ 86005, USA}

\author[0000-0001-7780-3352]{Michael C. Cushing}
\affil{Ritter Astrophysical Research Center, Department of Physics and Astronomy, University of Toledo, 2801 W. Bancroft St., Toledo, OH 43606, USA}

\author[0000-0002-7181-2554]{Robert A. Stiller}
\affil{Ritter Astrophysical Research Center, Department of Physics and Astronomy, University of Toledo, 2801 W. Bancroft St., Toledo, OH 43606, USA}

\author[0000-0002-4603-4834]{Jeffrey A. Munn}
\affil{United States Naval Observatory, Flagstaff Station, 10391 West Naval Observatory Rd., Flagstaff, AZ 86005, USA}

\author{Frederick J. Vrba}
\affil{United States Naval Observatory, Flagstaff Station, 10391 West Naval Observatory Rd., Flagstaff, AZ 86005, USA}

\author[0000-0002-3858-1205]{Justice Bruursema}
\affil{United States Naval Observatory, Flagstaff Station, 10391 West Naval Observatory Rd., Flagstaff, AZ 86005, USA}

\author{Stephen J. Williams}
\affil{United States Naval Observatory, Flagstaff Station, 10391 West Naval Observatory Rd., Flagstaff, AZ 86005, USA}

\author[0000-0003-2232-7664]{Michael C. Liu}
\affil{Institute for Astronomy, University of Hawaii at Manoa, Honolulu, HI 96822, USA}

\author[0009-0002-3936-8059]{Alexia Bravo}
\affil{United States Naval Observatory, Flagstaff Station, 10391 West Naval Observatory Rd., Flagstaff, AZ 86005, USA}

\author[0000-0001-6251-0573]{Jacqueline K. Faherty}
\affil{Department of Astrophysics, American Museum of Natural History, Central Park West at 79th St., New York, NY 10024, USA}

\author[0000-0003-4083-9962]{Austin Rothermich}
\affil{Department of Astrophysics, American Museum of Natural History, Central Park West at 79th St., New York, NY 10024, USA}

\author[0000-0002-2682-0790]{Emily Calamari}
\affil{Department of Astrophysics, American Museum of Natural History, Central Park West at 79th St., New York, NY 10024, USA}

\author[0000-0001-7896-5791]{Dan Caselden}
\affil{Department of Astrophysics, American Museum of Natural History, Central Park West at 79th St., New York, NY 10024, USA}

\author[0000-0003-4905-1370]{Martin Kabatnik}
\affil{Backyard Worlds: Planet 9, USA}

\author[0000-0003-4864-5484]{Arttu Sainio}
\affil{Backyard Worlds: Planet 9, USA}

\author[0000-0003-2235-761X]{Thomas P. Bickle}
\affil {School of Physical Sciences, The Open University, Milton Keynes, MK7 6AA, UK}
\affil {Backyard Worlds: Planet 9, USA}

\author{William Pendrill}
\affil{Backyard Worlds: Planet 9, USA}

\author[0000-0003-4714-3829]{Nikolaj Stevnbak Andersen}
\affil{Backyard Worlds: Planet 9, USA}

\author[0000-0001-5284-9231]{Melina Th{\'e}venot}
\affil{Backyard Worlds: Planet 9, USA}

\begin{abstract}

We have used the UKIRT Hemisphere Survey (UHS) combined with the UKIDSS Galactic Cluster Survey (GCS), the UKIDSS Galactic Plane Survey (GPS), and the CatWISE2020 catalog to search for new substellar members of the nearest open cluster to the Sun, the Hyades. Eight new substellar Hyades candidate members were identified and observed with the Gemini/GNIRS near-infrared spectrograph.  All eight objects are confirmed as brown dwarfs with spectral types ranging from L6 to T5, with two objects showing signs of spectral binarity and/or variability. A kinematic analysis demonstrates that all eight new discoveries likely belong to the Hyades cluster, with future radial velocity and parallax measurements needed to confirm their membership.   CWISE J042356.23$+$130414.3, with a spectral type of T5, would be the coldest (\teff$\approx$1100 K) and lowest-mass ($M$$\approx$30 \mjup) free-floating member of the Hyades yet discovered. We further find that high-probability substellar Hyades members from this work and previous studies have redder near-infrared colors than field-age brown dwarfs, potentially due to lower surface gravities and super-solar metallicities.
\end{abstract}

\keywords{Brown dwarfs; Open star clusters}

\section{Introduction}
\label{sec:intro}

As a cluster containing several naked-eye stars, the Hyades has been known since prehistory. The first instance of the Hyades cluster having been cataloged was likely in \cite{hodierna1654}, where it was labeled as a ``Luminosae'', or a region containing stars visible to the naked-eye. Compilations of Hyades members did not occur in earnest until the emergence of proper motion measurements, with the first recognition that some stars in the area of the Hyades may share similar proper motions (and thus be physically associated) noted in \cite{proctor1870}. Early compilations (e.g., \citealt{kustner1902, wirtz1902, kapteyn1904, weersman1904}) presented some of the first proper motion measurements in this area of the sky and catalogued dozens of members, with \cite{wirtz1902} likely showing the first map of the Hyades in a scientific journal.  It was not until the pioneering work of \cite{boss1908} where it was shown that the proper motion vectors of many proposed members were converging toward a single point in the sky, solidifying the physical association of many cluster members. Further studies continued to expand the number and properties of stellar Hyades members based on astrometric and photometric measurements  (e.g., \citealt{hertzsprung1921, vanrhijn1934, titus1940, ramberg1941, wilson1948, vanbueren1952, johnson1955, giclas1962, hanson1975, pels1975, upgren1977, oort1979, perryman1998, debruijne2001, roeser2011, reino2018, meingast2019, roeser2019}).  Such efforts to fill out the census of Hyades stars were complemented with targeted searches for fainter, lower-mass members (e.g., \citealt{vanmaanen1942, luyten1954, herbig1962, vanaltena1966, pesch1968, vanaltena1969, stauffer1982, zuckerman1987, leggett1989, bryja1992, reid1992, reid1993, bryja1994, leggett1994, stauffer1994, stauffer1995, reid1997, harris1999, gizis1999, reid1999, reid2000, dobbie2002, bannister2007, bouvier2008, hogan2008, goldman2013, casewell2014, lodieu2014, perez2017, perez2018, melnikov2018, lodieu2019, zhang2021, schneider2022}).  

Brown dwarfs, substellar objects that have masses below the hydrogen burning minimum mass ($M$ $\lesssim$ 0.075 $M_{\sun}$) and cannot sustain hydrogen fusion in their cores \citep{kumar1963, hayashi1963}, have been historically challenging to find in the Hyades. The main obstacles that arise when investigating the Hyades population below the stellar boundary are the cluster's large spatial extent on the sky and the intrinsic faintness of substellar objects at the distance of the Hyades ($\sim$47 pc; \citealt{gaia2021}. Despite these hurdles, several substellar members of the Hyades have been identified, either by focusing on an area smaller than the full extent of the cluster core (e.g., \citealt{bouvier2008, hogan2008}), conducting shallow (but wide) investigations that detected some of the nearest substellar members (e.g., \citealt{perez2017, perez2018}), performing high-contrast imaging to detect substellar companions to known stellar members \citep{kuzuhara2022, franson2023}, or complete serendipity (e.g., \citealt{schneider2017, schneider2023}).

Any such discoveries, however, hold incredible value as benchmark systems.  As one of the most well-studied clusters, the Hyades has an established age ($\sim$650 Myr; \citealt{lebreton2001, degennaro2009, gossage2018, martin2018, lodieu2019, lodieu2020}). Models that include rotation predict a slightly older age for the Hyades cluster (750$\pm$100 Myr; \citealt{brandt2015a, brandt2015b}), however these ages are in tension with previous estimates and more recent determinations using the lithium depletion boundary (e.g., \citealt{martin2018, lodieu2020}) and white dwarfs \citep{lodieu2019}.   The Hyades also has a well-determined metallicity ([Fe/H]$\sim$0.15 dex; \citealt{cummings2017}).  Such properties are exceptionally difficult to determine for solitary brown dwarfs not belonging to a known association.

While the spatial extent of the Hyades presents a challenge to its exploration, the proper motion of the cluster ($\sim$100 mas yr$^{-1}$) provides a significant advantage for distinguishing candidate substellar members from the myriad of background sources. Using new proper motion measurements, \cite{schneider2022} performed a large area search for new candidate substellar members of the Hyades using the UKIRT Hemisphere Survey (UHS; \citealt{dye2018}), which reaches $\sim$4 magnitudes deeper than the Two Micron All-Sky Survey (2MASS; \citealt{skrutskie2006}).  Candidates were found by combining UHS with the CatWISE2020 catalog \citep{marocco2021}, both of which cover the majority of the spatial extent of the Hyades. 

In this work, we have expanded the search of \cite{schneider2022}, probing deeper in the cluster and to colder temperatures, identifying eight candidates with colors consistent with known field late-L or T dwarfs and motions consistent with Hyades membership.  In Section \ref{sec:targets} we describe the target selection for these new substellar Hyades candidates.  In Section \ref{sec:obs} we describe our follow-up spectroscopic observations.  In Section \ref{sec:anal} we analyze the characteristics of each new candidate, including spectral types, distance estimates, Hyades membership, and physical properties.  Finally in Section \ref{sec:disc} we discuss the near-infrared colors of the current census of substellar Hyades members compared to those of the field-age brown dwarf population, and summarize our results in Section \ref{sec:summary}.

\section{Target Selection}
\label{sec:targets}

We use a similar candidate selection strategy as in \cite{schneider2022} with a few key differences.  \cite{schneider2022} searched for substellar Hyades candidates using the UHS and CatWISE2020 catalogs, limiting new candidates to those having $J$-band magnitudes brighter than 17.5 mag.  This limit was imposed such that a spectrum with a reasonable signal-to-noise (S/N)  could be obtained with the SpeX spectrograph on the 3m IRTF telescope.  For this current search, we have increased the $J$-band magnitude limit to 18 mag.  This magnitude limit was chosen to focus on the brightest new candidates.  Deeper searches for candidates are feasible, as UHS has a  $J$-band 5$\sigma$ limiting magnitude of 19.6 \citep{dye2018}.  We also include detections in previous UKIDSS surveys \citep{lawrence2007} combined with UHS to determine proper motions of new candidates, extending the time baseline for proper motion measurements from $\sim$5 to $\sim$10 years.  Following the methods described in \cite{schneider2023}, we recalibrate the position of each UKIRT detection for each source using the Gaia DR3 \citep{gaia2023} reference frame.  

\begin{figure}
\plotone{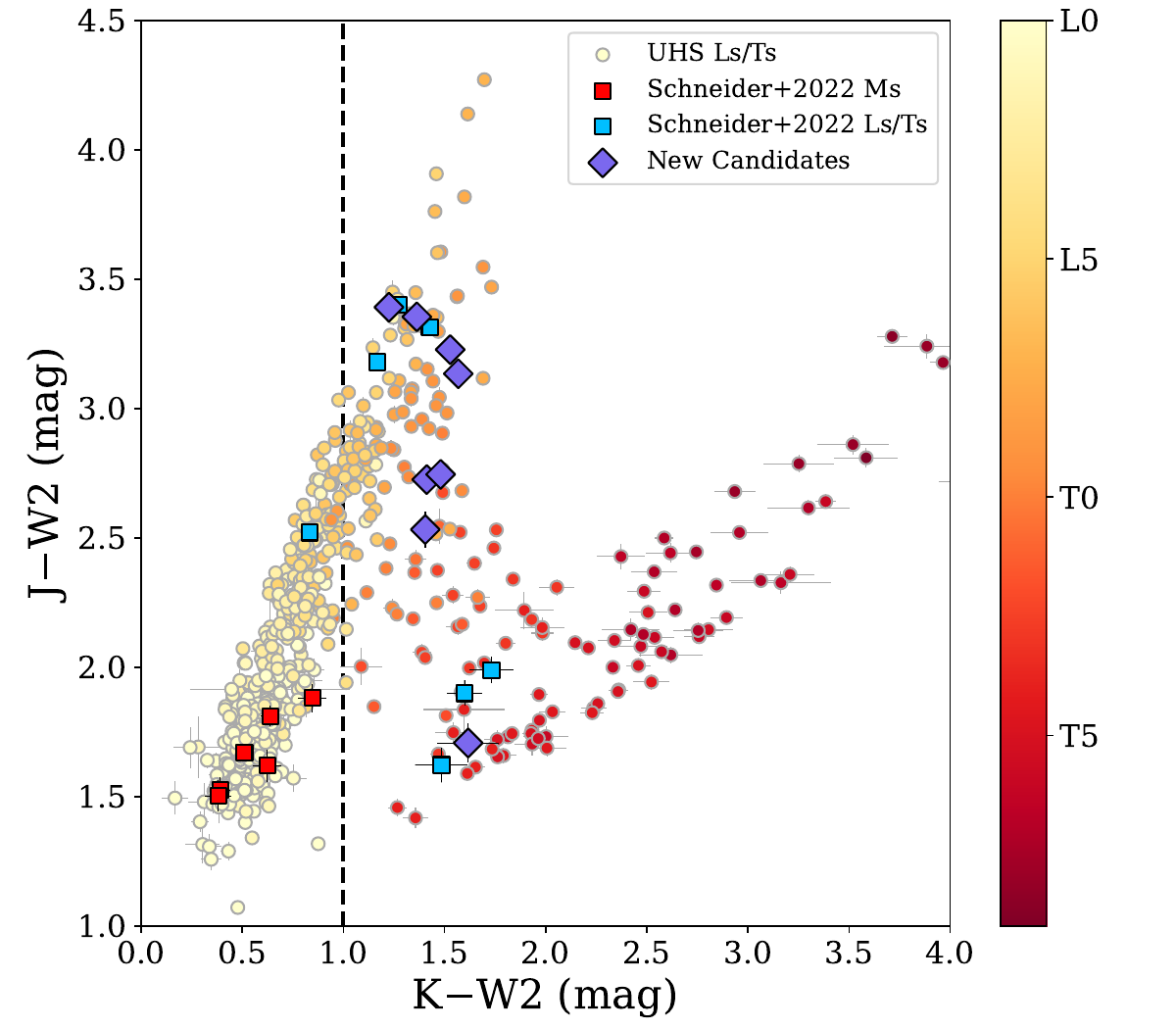}
\caption{$J-$W2 versus $K-$W2 colors for known L and T dwarfs from the UKIRT Hemisphere Survey \citep{schneider2023} compared to discoveries from \cite{schneider2022} and new candidates presented in this work.  Note that the color cut requiring new candidates have $K-$W2 colors greater than 1.0 mag effectively eliminates the late-M type objects that contaminated the \cite{schneider2022} sample. }  
\label{fig:colors}
\end{figure}

Following \cite{schneider2022}, we have limited our search to objects that are within 18 pc of the Hyades cluster center, which should include all bound and halo cluster members \citep{lodieu2019}. Candidates were chosen to also have colors consistent with late-L and T dwarfs.  Specifically, we chose objects with $J-W2$ colors greater than 1.5 mag and $K-W2$ colors greater than 1.0 mag.  The first color requirement selects red, late-type ($\gtrsim$M7) objects, while the second color criterion effectively filters out late-M dwarfs, such as those that contaminated the \cite{schneider2022} sample.  Our color selection criteria is illustrated in Figure \ref{fig:colors}.  

\begin{figure*}
\plotone{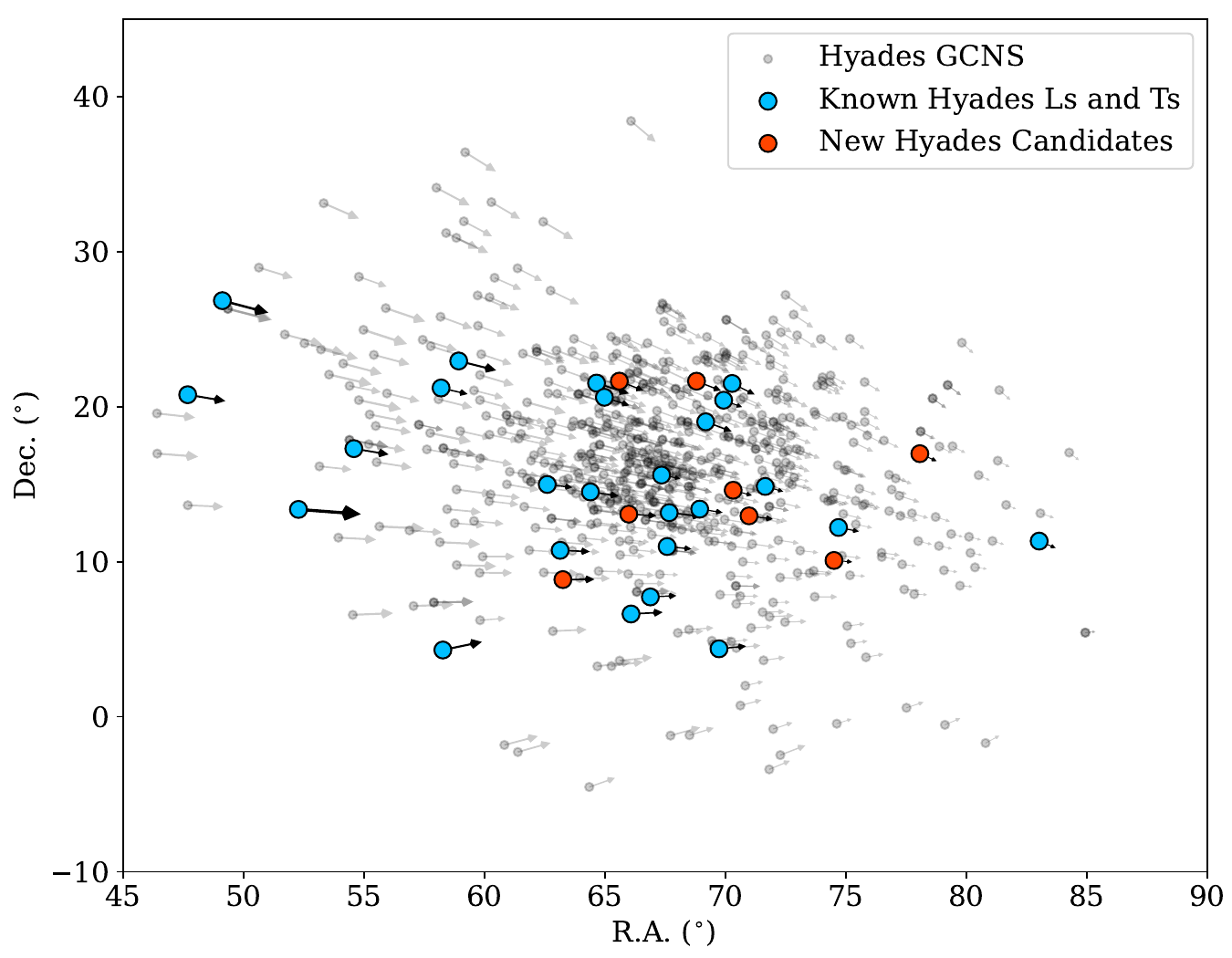}
\caption{The positions and proper motion vectors of our new substellar Hyades candidates (red) and known L and T type candidate members of the Hyades (blue; \citealt{schneider2022}) compared to all known Hyades members (grey) within the halo radius (18 pc) from the Gaia Catalog of Nearby Stars (GCNS; \citealt{gaia2021}).}  
\label{fig:plot1}
\end{figure*}

To ensure each candidate has proper motion components consistent with known Hyades members from the Gaia Catalog of Nearby Stars (GCNS; \citealt{gaia2021}), we verified each source is a kinematic match to the Hyades central cluster using the BANYAN $\Sigma$ classifier \citep{gagne2018} and a convergent point analysis (e.g., \citealt{schneider2022}). Specifically, we required that the BANYAN $\Sigma$ probability of Hyades membership using positions and measured proper motions to be greater than 0\%. We chose this conservative value to ensure no potential candidates are missed, however we note that most of our final candidates have BANYAN $\Sigma$ probabilities $\geq$80\%.  For the convergent point analysis, we required the proper motion angle ($\theta_{\mu}$) and the convergent point angle ($\theta_{\rm cp}$), defined as the angle between a line pointing north and a line from each candidate pointed to the convergent point, to be within 7\degr\ of each other ($\sim$3$\sigma$).  This search revealed several new substellar Hyades candidate members. This work focuses on the eight brightest new candidates ($J\leq$18 mag).  The positions and proper motion vectors of these candidates compared to known Hyades members are shown in Figure \ref{fig:plot1}.

Of the eight candidates discussed in this work, seven were imaged by previous UKIRT Surveys, either in the UKIDSS Galactic Plane Survey (GPS; \citealt{lucas2008}) or Galactic Cluster Survey (GCS; \citealt{lawrence2007}).  Two of the candidates were found in both the GCS and GPS (CWISE J042222.17$+$213900.5 and CWISE J043511.26$+$213846.3). CWISE J044354.22$+$125736.7 has an available UKIDSS GCS $K$-band image, but has no matching entries in the UKIDSS GCS catalog.  For this image, we use the \texttt{imcore} routine from the \texttt{CASUTOOLS} package\footnote{http://casu.ast.cam.ac.uk/surveys-projects/software-release} \citep{irwin2004} to extract source positions and photometry.  We recalibrate these extracted positions in the same method as the other survey images and provide the position of this source from the UKIDSS GCS $K$-band image in Table \ref{tab:astro}.  Offsets between catalog positions and our recalibrations for all eight candidates were between 16 mas and 163 mas, with an average offset of 92 mas. 

\begin{deluxetable*}{cccccccc}
\label{tab:astro}
\tablecaption{Astrometric Measurements\tablenotemark{a}}
\tablehead{
\colhead{CWISE} & \colhead{R.A.} & \colhead{R.A. err} & \colhead{Dec.} & \colhead{Dec err} & \colhead{Epoch} & \colhead{Survey} \\
\colhead{Name} & \colhead{($\degr$)} & \colhead{(mas)} & \colhead{($\degr$)} & \colhead{(mas)} & \colhead{(year)} & \colhead{}  }
\startdata
J041259.89$+$085049.6 & 63.249233276 & 24.2843 & 8.847085790 & 25.5046 & 2006.0119 & GCS ($K$) \\ 
\dots & 63.249487662 & 23.7192 & 8.847089852 & 24.7934 & 2013.1109 & UHS ($J$) \\ 
\dots & 63.249729257 & 13.3274 & 8.847090340 & 13.7931 & 2018.8528 & UHS ($K$) \\ 
J042222.17$+$213900.5 & 65.592092834 & 23.6696 & 21.650238453 & 21.1248 & 2005.7772 & GCS ($K$) \\ 
\dots & 65.592082447 & 27.6446 & 21.650231543 & 28.5865 & 2005.9108 & GPS ($J$) \\ 
\dots & 65.592096152 & 19.8400 & 21.650230210 & 19.6598 & 2005.9108 & GPS ($H$) \\ 
\dots & 65.592088999 & 21.2632 & 21.650231288 & 22.7655 & 2005.9108 & GPS ($K$) \\ 
\dots & 65.592338488 & 19.9523 & 21.650140765 & 19.2822 & 2013.9792 & UHS ($J$) \\ 
\dots & 65.592478643 & 13.9178 & 21.650092527 & 11.0290 & 2018.0148 & UHS ($K$) \\ 
J042356.23$+$130414.3 & 65.983999636 & 95.7351 & 13.070798587 & 95.0150 & 2005.7798 & GCS ($K$) \\ 
\dots & 65.984283283 & 13.7973 & 13.070763288 & 13.7151 & 2013.9002 & UHS ($J$) \\ 
\dots & 65.984426861 & 46.6846 & 13.070765354 & 46.7912 & 2018.1185 & UHS ($K$) \\ 
J043511.26$+$213846.3 & 68.796669723 & 24.3706 & 21.646361000 & 22.7210 & 2005.7773 & GCS ($K$) \\ 
\dots & 68.796659215 & 23.7114 & 21.646367350 & 23.2133 & 2005.8866 & GPS ($J$) \\ 
\dots & 68.796661991 & 20.1629 & 21.646368124 & 18.9891 & 2005.8866 & GPS ($H$) \\ 
\dots & 68.796652499 & 32.7252 & 21.646357953 & 36.6166 & 2005.8866 & GPS ($K$) \\ 
\dots & 68.796951296 & 13.3308 & 21.646258961 & 13.0415 & 2015.0361 & UHS ($J$) \\ 
\dots & 68.797045604 & 13.8996 & 21.646234299 & 14.6700 & 2017.9273 & UHS ($K$) \\ 
\dots & 68.797077425 & 16.0486 & 21.646206952 & 16.8108 & 2018.7874 & UHS ($K$) \\ 
J044116.07$+$143645.4 & 70.316806464 & 66.7943 & 14.612599107 & 67.3614 & 2005.7799 & GCS ($K$) \\ 
\dots & 70.317022493 & 40.0415 & 14.612538522 & 39.5753 & 2015.0302 & UHS ($J$) \\ 
\dots & 70.317071886 & 36.3958 & 14.612505110 & 34.7169 & 2016.5561 & UHS ($J$) \\ 
\dots & 70.317067518 & 34.4074 & 14.612531056 & 34.6010 & 2016.9034 & UHS ($J$) \\ 
\dots & 70.317091733 & 21.2303 & 14.612524547 & 21.6415 & 2017.8672 & UHS ($K$) \\ 
J044354.22$+$125736.7 & 70.975750977 & 19.6853 & 12.960214359 & 19.6748 & 2009.6828 & GCS ($K$) \\ 
\dots & 70.975908997 & 17.5431 & 12.960198900 & 17.4367 & 2014.9459 & UHS ($J$) \\ 
\dots & 70.975998969 & 19.6072 & 12.960185199 & 18.4786 & 2017.8670 & UHS ($K$) \\ 
\dots & 70.975998431 & 15.1848 & 12.960177469 & 14.8130 & 2018.0502 & UHS ($K$) \\ 
J045800.69$+$100456.9 & 74.502673652 & 44.1223 & 10.082517681 & 43.8587 & 2005.8727 & GCS ($K$) \\ 
\dots & 74.502889253 & 18.1274 & 10.082491683 & 17.4784 & 2014.9405 & UHS ($J$) \\ 
\dots & 74.502943664 & 21.2104 & 10.082496415 & 20.9111 & 2017.8427 & UHS ($K$) \\ 
J051215.86$+$165818.0 & 78.066057176 & 18.8927 & 16.971622726 & 17.5606 & 2013.0839 & UHS ($J$) \\ 
\dots & 78.066171159 & 17.2959 & 16.971585306 & 17.9962 & 2018.0065 & UHS ($K$) \\ 
\dots & 78.066150641 & 24.5892 & 16.971566542 & 25.461 & 2018.0175 & UHS ($K$) \\ 
\enddata
\tablenotetext{a}{All positions and uncertainties are determined by astrometrically recalibrating UHS and UKIDSS positions using the Gaia DR3 reference frame.}
\end{deluxetable*}

The proper motions and photometric measurements of the eight new candidates are listed in Table \ref{tab:props}.  We include photometry from CatWISE2020 \citep{marocco2021}, UHS \citep{dye2018} and Pan-STARRS (PS1) DR2 \citep{chambers2016, magnier2020}.  Several of these sources were independently identified as candidate brown dwarfs through the Backyard Worlds: Planet 9 citizen science project \citep{kuchner2017}.  These volunteers are credited as co-discoverers and noted in Table \ref{tab:props}.

\begin{longrotatetable} 
\begin{deluxetable*}{lrrccccccccccc}
\label{tab:props}
\tablecaption{New Substellar Hyades Candidates}
\tablehead{
\colhead{CWISE} & \colhead{$\mu_{\alpha}$} & \colhead{$\mu_{\delta}$} & \colhead{$z_{\rm PS1}$} & \colhead{$y_{\rm PS1}$} & \colhead{$J_{\rm UHS}$} & \colhead{$K_{\rm UHS}$} & \colhead{W1} & \colhead{W2}  \\
\colhead{Name} & \colhead{(mas yr$^{-1}$)} & \colhead{(mas yr$^{-1}$)} & \colhead{(mag)} & \colhead{(mag)} & \colhead{(mag)} & \colhead{(mag)} & \colhead{(mag)} & \colhead{(mag)}}
\startdata
J041259.89$+$085049.6\tablenotemark{a} & 138.20$\pm$2.58  &   1.21$\pm$2.69  & 21.223$\pm$0.017 & 19.706$\pm$0.153 & 17.530$\pm$0.044 & 16.216$\pm$0.040 & 15.460$\pm$0.022 & 14.804$\pm$0.025 \\
J042222.17$+$213900.5\tablenotemark{b} & 106.13$\pm$1.87  & -41.57$\pm$1.61  & 20.939$\pm$0.152 & 20.028$\pm$0.279 & 17.712$\pm$0.042 & 15.545$\pm$0.023 & 14.709$\pm$0.017 & 14.320$\pm$0.019 \\
J042356.23$+$130414.3 & 120.96$\pm$8.36 &  -7.19$\pm$8.34 & \dots & 20.145$\pm$0.090 & 17.651$\pm$0.032 & 17.560$\pm$0.136 & 17.138$\pm$0.062 & 15.943$\pm$0.067 \\
J043511.26$+$213846.3\tablenotemark{c} & 106.80$\pm$1.55  & -41.49$\pm$1.54  & 20.864$\pm$0.043 & 20.076$\pm$0.022 & 17.894$\pm$0.043 & 15.902$\pm$0.032 & 14.998$\pm$0.019 & 14.539$\pm$0.023 \\
J044116.07$+$143645.4 &  82.64$\pm$6.18  & -22.91$\pm$6.24  & 20.733$\pm$0.047 & 20.149$\pm$0.190 & 17.932$\pm$0.056 & 16.804$\pm$0.061 & 16.078$\pm$0.029 & 15.399$\pm$0.042 \\
J044354.22$+$125736.7\tablenotemark{d} &  104.76$\pm$3.57  & -14.84$\pm$3.51  & 20.927$\pm$0.173 & 20.065$\pm$0.049 & 17.981$\pm$0.040 & 16.281$\pm$0.038 & 15.343$\pm$0.020 & 14.753$\pm$0.025 \\
J045800.69$+$100456.9\tablenotemark{e} &  80.05$\pm$4.62  &  -6.45$\pm$4.55  & \dots & \dots & 17.901$\pm$0.043 & 16.636$\pm$0.060 & 15.859$\pm$0.027 & 15.155$\pm$0.034 \\
J051215.86$+$165818.0 &  73.92$\pm$6.47  & -32.84$\pm$6.49  & \dots & 20.007$\pm$0.198 & 17.698$\pm$0.039 & 16.131$\pm$0.042 & 15.110$\pm$0.019 & 14.563$\pm$0.022 \\
\enddata
\tablenotetext{a}{CWISE J041259.89+085049.6 was independently discovered by citizen scientists Melina Th{\'e}venot, Arttu Sainio, Sam Goodman, and Martin Kabatnik.}
\tablenotetext{b}{CWISE J042222.17+213900.5 was independently discovered by citizen scientists Nikolaj Stevnbak Andersen, William Pendrill, and Tom Bickle.}
\tablenotetext{c}{CWISE J043511.26+213846.3 was independently discovered by citizen scientists Dan Caselden and Tom Bickle.}
\tablenotetext{d}{CWISE J044354.22+125736.7 was independently discovered by citizen scientists Dan Caselden, Melina Th{\'e}venot, Sam Goodman, and William Pendrill.}
\tablenotetext{e}{CWISE J045800.69+100456.9 was independently discovered by citizen scientists Arttu Sainio and Martin Kabatnik.}
\end{deluxetable*}
\end{longrotatetable}

\section{Observations}
\label{sec:obs}

All eight substellar Hyades candidates were observed with the Gemini Near-InfraRed Spectrograph (GNIRS; \citealt{elias2006}) at the Gemini North Telescope in November 2023 (PID: GN-2023B-Q-316).  GNIRS was operated in the non-AO cross-dispersed mode with the 1\arcsec\ slit and the 32 l/mm grating (R=500, 0.8-2.5 $\mu$m). Each target was dithered between two positions along the slit separated by 3\arcsec\ in an ABBA pattern. A summary of the observations is given in Table \ref{tab:obs}.  This table includes the total exposure time and the S/N per pixel at the $J$-band peak between 1.27 and 1.29 $\mu$m. 

\begin{figure*}
\plotone{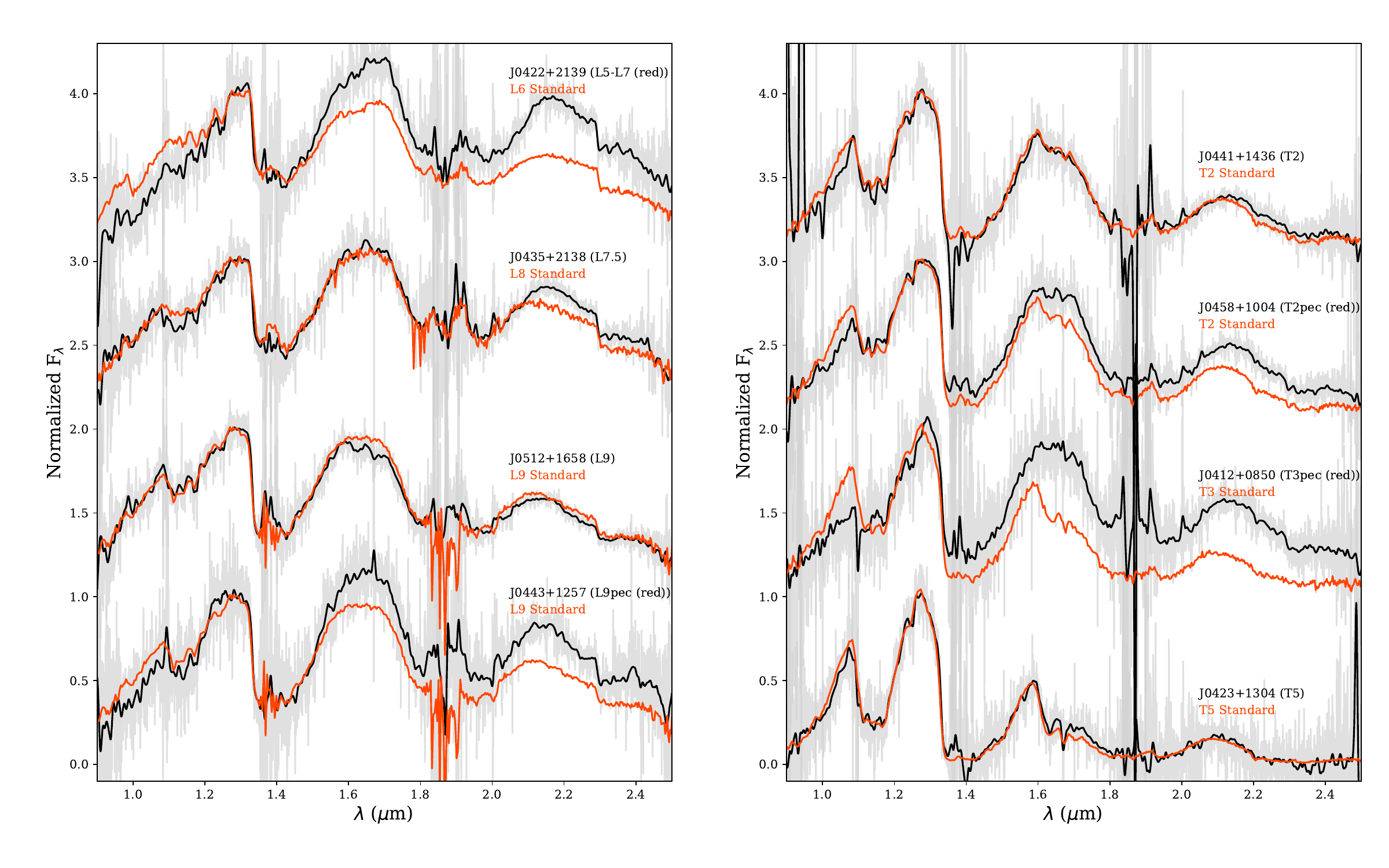}
\caption{Gemini/GNIRS spectra in full-resolution (grey) and smoothed (black) compared to spectral standards (red-orange).  The spectra are normalized between 1.27 and 1.29 $\mu$m and offset by integer values for clarity.  The spectral standards are: 2MASSI J1010148$-$040649 (L6; \citealt{reid2006}), 2MASSW J1632291$+$190441 (L8; \citealt{burgasser2007}), DENIS-P J0255-4700 (L9; \citealt{burgasser2006}), SDSSp J125453.90$-$012247.4 (T2; \citealt{burgasser2004}); 2MASS J12095613$-$1004008 (T3; \citealt{burgasser2004}); 2MASS J15031961$+$2525196 (T5; \citealt{burgasser2004}).}  
\label{fig:spectra}
\end{figure*}

All spectra were reduced with the open-source Python package \texttt{PypeIt} \citep{prochaska2020, prochaska2023}. \texttt{PypeIt} performs flat-fielding, wavelength calibration, flux calibration, and telluric corrections.  The reduced spectra are shown in Figure \ref{fig:spectra}. 

\begin{deluxetable*}{lcccccc}
\label{tab:obs}
\tablecaption{GNIRS Observations}
\tablehead{
\colhead{CWISE} & \colhead{Obs.\ Date} & \colhead{Total Exp.~Time} & \colhead{A0 Star\tablenotemark{a}} & \colhead{Spec.} & \colhead{(S/N)$_J$\tablenotemark{b}} \\
\colhead{Name} & \colhead{(UT)} & \colhead{(s)} & \colhead{} & \colhead{Type} & \colhead{}  }
\startdata
J041259.89$+$085049.6 & 2023 Nov 14 & 2400 & HIP 22923 & T3pec (red) & 11 \\ 
J042222.17$+$213900.5 & 2023 Nov 14 & 2880 & HIP 16095 & L5--L7 (red) & 16 \\ 
J042356.23$+$130414.3 & 2023 Nov 15 & 2400 & HIP 16095 & T5 & 15 \\ 
J043511.26$+$213846.3 & 2023 Nov 19 & 4320 & HIP 17453 & L7.5 & 16 \\ 
J044116.07$+$143645.4 & 2023 Nov 20 & 4320 & HIP 16095 & T2 & 15 \\ 
J044354.22$+$125736.7 & 2023 Nov 15 & 2400 & HIP 28686 & L9pec (red) & 9 \\ 
J045800.69$+$100456.9 & 2023 Nov 16 & 1440 & HIP 16095 & T2pec (red) & 12 \\ 
J051215.86$+$165818.0 & 2023 Nov 19 & 2400 & HIP 29371 & L9 & 18 \\ 
\enddata
\tablenotetext{a}{The A0 star used for telluric correction.}
\tablenotetext{b}{Signal-to-noise determined at the $J$-band peak.}
\end{deluxetable*}

\section{Analysis}
\label{sec:anal}

\subsection{Spectral Types}
\label{sec:spts}

Spectral types for each source were determined by comparing the observed spectra to L and T dwarf near-infrared spectral standards from \cite{burgasser2006} and \cite{kirkpatrick2010}.  The best-fit spectral types were determined by finding the minimum $\chi$$^2$ values comparing the $J$-band portion of each spectrum (1.0--1.35 $\mu$m) to each spectral standard. The best-fit types are given in Table \ref{tab:obs} and comparisons to the best-fitting standards are shown in Figure \ref{fig:spectra}. Additional notes on individual spectra are given in the following subsections. 

\subsubsection{CWISE J043511.26$+$213846.3, CWISE J051215.86$+$165818.0, CWISE J044116.07$+$143645.4, and CWISE J042356.23$+$130414.3}
All four of these objects show reasonably good fits to their respective best-matching standards. Of these, CWISE J042356.23$+$130414.3 is particularly noteworthy with a T5 spectral type.  If confirmed as a Hyades member, CWISE J042356.23$+$130414.3 would have the latest spectral type amongst Hyades free-floating Hyades members. 

\subsubsection{CWISE J042222.17$+$213900.5}
\begin{figure}
\plotone{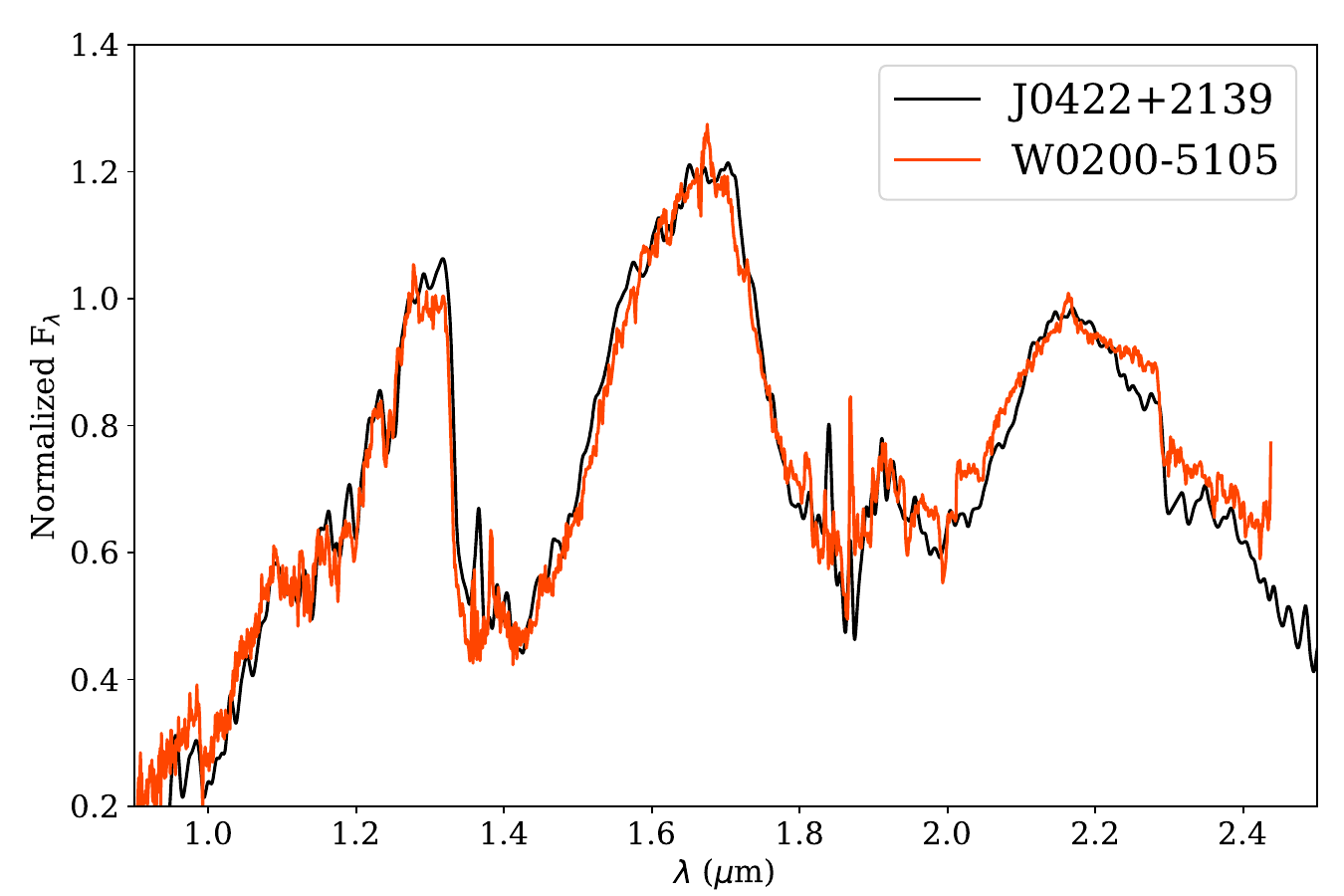}
\caption{Comparison of the near-infrared spectrum of CWISE J042222.17$+$213900.5 to the likely AB Dor member WISEA J020047.29$-$510521.4 \citep{schneider2017}.}  
\label{fig:redl}
\end{figure}

CWISE J042222.17$+$213900.5 is unusually red compared to the L6 spectral standard.  Figure \ref{fig:redl} shows that this object is a very good match spectroscopically to WISEA J020047.29$-$510521.4 \citep{schneider2017}.  \cite{schneider2017} suggested WISEA J020047.29$-$510521.4 as a probable AB Dor member ($\sim$150 Myr; \citealt{bell2015}).  Using updated astrometry of this source from \cite{kirkpatrick2021}, which includes proper motion components $\sim$5$\times$ more precise than the proper motion previously used to evaluate moving group membership and a measured parallax, we find a 99.5\% probability of belonging to AB Dor using BANYAN $\Sigma$ \citep{gagne2018}.  

Unusually red colors are often associated with low-surface gravity, though there are exceptions (e.g., \citealt{marocco2014}).  We evaluated the surface gravity of CWISE J042222.17$+$213900.5 using the \cite{allers2013} spectral indices that are applicable to objects with spectral types of L6 (FeH$_J$ and $H$-cont). We find values of 1.17 and 0.87 for FeH$_J$ and $H$-cont, respectively, which indicates a normal field gravity according to \cite{allers2013}.  We also evaluated the H$_2$(K) index \citep{canty2013, schneider2014}, and again find a value (1.12) consistent with the field population.  This suggests that while CWISE J042222.17$+$213900.5 is unusually red, it does not have a surface gravity low enough to distinguish it from the field population. WISEA J020047.29$-$510521.4, on the other hand, has FeH$_J$, $H$-cont, and H$_2$(K) values (1.06, 0.90, and 1.07, respectively) that suggest low to intermediate surface gravity, as expected for an AB Dor member.  Thus, while the overall spectral shape of CWISE J042222.17$+$213900.5 and WISEA J020047.29$-$510521.4 are similar, the gravity sensitive regions show that the surface gravities of these two sources are distinct.  

\subsubsection{CWISE J044354.22$+$125736.7}
CWISE J044354.22$+$125736.7 most closely matches the L9 standard at $J$, but is not a great fit elsewhere.  We note that this is the lowest-S/N spectrum presented in this work, and a higher-S/N spectrum may be needed for a more accurate classification.  

\subsubsection{CWISE J041259.89$+$085049.6 and CWISE J045800.69$+$100456.9}
\begin{figure*}
\plotone{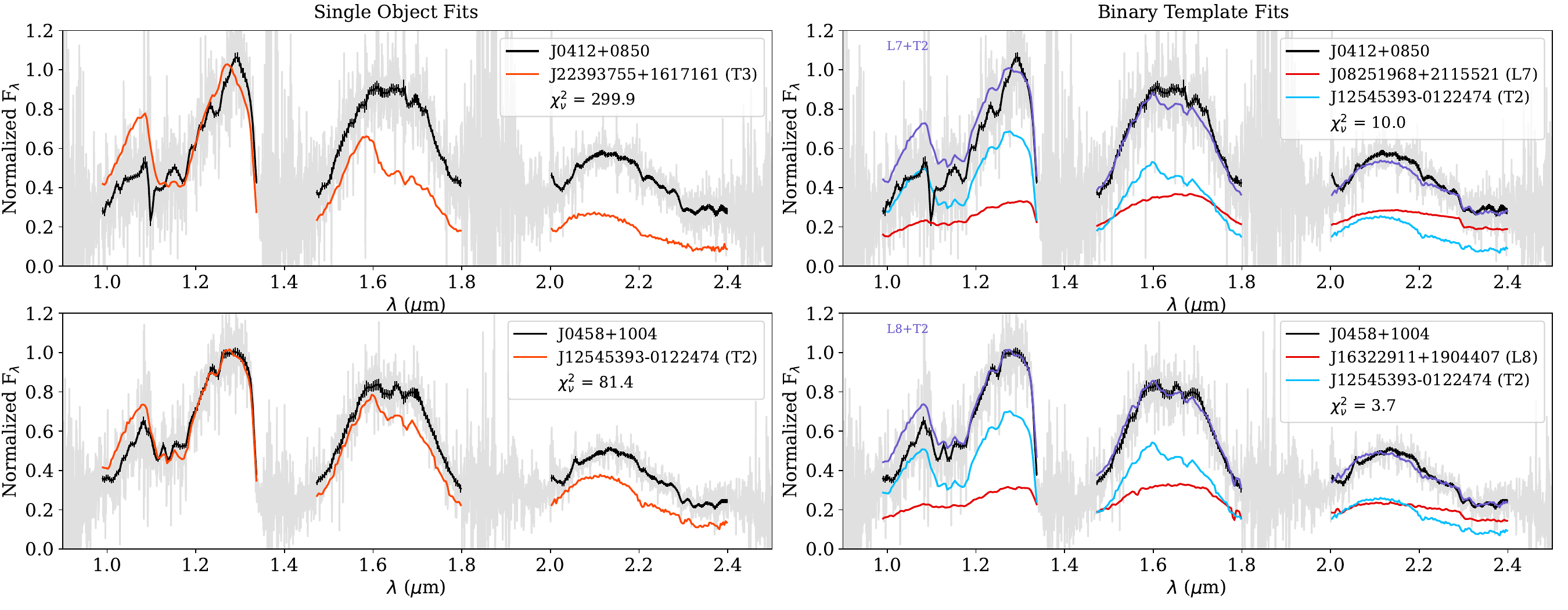}
\caption{Spectral binary fits for CWISE J041259.89$+$085049.6 and CWISE J045800.69$+$100456.9.  The individual components used in the fits are 2MASSI J0825196$+$211552 (L7; \citealt{kirkpatrick2000, burgasser2010}), 2MASSW J1632291+190441 (L8; \citealt{kirkpatrick1999, burgasser2007}), SDSS J125453.90$-$012247.5 (T2; \citealt{leggett2000, burgasser2004}), and WISEPC J223937.55$+$161716.2 (T3; \citealt{kirkpatrick2011}). }  
\label{fig:binfit}
\end{figure*}

The single standard fits to CWISE J041259.89$+$085049.6 and CWISE J045800.69$+$100456.9 are relatively poor, with the best-fitting standards providing decent fits at $J$, but not aligning well beyond $\sim$1.4 $\mu$m. Using the methods outlined in \cite{bravo2023}, we looked for spectral binary templates that better fit the observed spectra for each of these objects than the spectral standards.  The best-fitting binary templates for each of CWISE J041259.89$+$085049.6 (L7+T2) and CWISE J045800.69$+$100456.9 (L8+T2) are shown in Figure \ref{fig:binfit}.  The binary template fits are clearly better matches to the observed spectra for both objects, as reflected in their much lower $\chi^2_{\nu}$ values.  

However, spectral indices potentially point to variability due to a patchy atmosphere as an alternative explanation. A brown dwarf atmosphere with patches of thick, cold clouds and thinner areas that reveal the deeper, hotter layers can have a spectrum that looks very similar to a spectral binary because it includes the contributions of multiple layers in an atmosphere at different temperatures (e.g., 2MASS J21392676$+$0220226 \citealt{burgasser2010, radigan2012}). Using the methods and indices defined in \cite{burgasser2002, burgasser2006, burgasser2010} and \cite{bardalez2014}, we compared the measured index values for each of these two sources to the binary index regions from \cite{burgasser2010} and variability regions from \cite{ashraf2022}.  Both CWISE J041259.89$+$085049.6 and CWISE J045800.69$+$100456.9 only satisfy a single binary index criterion from \cite{burgasser2010} (H$_2$O$-J$/CH$_4$-$K$ versus spectral type).  Therefore the index values of these two sources do not even satisfy the threshold to be labeled ``weak'' spectral binary candidates, which is applied to those objects falling in at least two of the spectral binary index defined regions.  However, both objects satisfy all eleven of the variability index criteria defined in \cite{ashraf2022}.  Thus, variability may be the cause of the spectral peculiarity of these objects.  Photometric and/or spectroscopic monitoring, as well as high resolution imaging, may help to identify the causes of these spectral peculiarities. 

\subsection{Distances}
\label{sec:astrometry}
Photometric distances are determined using the spectral type vs.\ absolute magnitude relations from \cite{schneider2023} for L, T, and Y dwarfs with UHS photometry and measured parallaxes.  The distances given in Table \ref{tab:member} are the weighted average of the $J$- and $K$-band distances, which are determined considering photometric uncertainties, a $\pm$0.5 spectral subtype uncertainty, and the RMS scatter of the polynomial fits.  The exceptions are the variability candidates (CWISE J041259.89$+$085049.6 and CWISE J045800.69$+$100456.9), the very red CWISE J042222.17$+$213900.5, and the peculiar CWISE J044354.22$+$125736.7, for which we use $\pm$1 spectral subtype uncertainties. 

\subsection{Hyades Membership}
\label{sec:membership}
To evaluate the likelihood of Hyades membership for each candidate, we use the BANYAN $\Sigma$ \citep{gagne2018} moving group probability calculator as well as a convergent point analysis.  Table \ref{tab:member} summarizes the membership evaluation results.  For convergent point assessment, we use the convergent point from \cite{madsen2002} and follow the methods in \cite{hogan2008} and \cite{schneider2022}.  As seen in the table, the predicted distances from BANYAN $\Sigma$ and the convergent point analysis are generally consistent with our photometric distances estimates.  Table \ref{tab:member} also shows that the measured convergent point angles ($\theta_{\rm cp}$) and proper motion angles ($\theta_{\mu}$) are consistent to within 4$\degr$ ($\sim$2$\sigma$).  The table further provides two BANYAN $\Sigma$ probabilities, the first using only positions and proper motions, while the second includes photometric distance estimates. We include the BANYAN $\Sigma$ probability without using the photometric distances in case the photometric distance estimate is inaccurate, as would be the case for spectral binaries.  However, we note that all BANYAN $\Sigma$ probabilities increase when photometric distances are included and are greater than 90\% for each candidate member.  Considering these high BANYAN $\Sigma$ probabilities, the consistency between photometric distance estimates and distance predictions if true Hyades members, as well as the small differences between $\theta_{\rm cp}$ and $\theta_{\mu}$, we suggest that each new candidate is a likely Hyades member.  Directly measured distances and radial velocities are still needed, however, to fully confirm Hyades membership.

The success rate of these observations for recovering Hyades candidates is high.  We attribute this success rate to several factors: 1) the relatively localized area of the sky we limit to around the core of the Hyades cluster, 2) the lessons learned from \cite{schneider2022} regarding effective color cuts to filter out non-brown dwarf background contaminants, and 3) the pre-selection based on BANYAN $\Sigma$ probabilities and convergent point restrictions.   

\begin{deluxetable*}{lcrccrrccc}
\label{tab:member}
\tablecaption{New Hyades Candidate Membership Details}
\tablehead{\colhead{CWISE} & \colhead{Spec.} & \colhead{dist$_{\rm phot}$\tablenotemark{$\dagger$}} & \colhead{dist$_{\rm cp}$\tablenotemark{$\dagger$}} & \colhead{dist$_{\rm BANYAN}$\tablenotemark{$\dagger$}} & \colhead{$\theta_{\mu}$\tablenotemark{$\ddagger$}} & \colhead{$\theta_{\rm cp}$\tablenotemark{$\ddagger$}} & \colhead{BANYAN\tablenotemark{a}}  & \colhead{BANYAN\tablenotemark{b}} \\
\colhead{Name} & \colhead{Type} & \colhead{(pc)} & \colhead{(pc)} & \colhead{(pc)} & \colhead{($\degr$)} & \colhead{($\degr$)} & \colhead{(\%)} & \colhead{(\%)}  }
\startdata
J041259.89$+$085049.6 & T3pec (red) & 31$\pm$8 & 39.4 & 39.1 &  89.5 &  90.8 & 95.9 & 98.2 \\  
J042222.17$+$213900.5 & L5-L7 (red) & 54$\pm$7 & 48.1 & 47.6 & 111.4 & 110.9 & 95.4 & 97.6 \\
J042356.23$+$130414.3 & T5          & 38$\pm$3 & 42.2 & 40.7 &  94.7 &  98.2 & 91.4 & 98.3 \\
J043511.26$+$213846.3 & L7.5        & 51$\pm$6 & 44.3 & 44.6 & 111.2 & 113.9 & 91.1 & 95.0 \\
J044116.07$+$143645.4 & T2          & 42$\pm$6 & 53.0 & 51.1 & 106.2 & 103.3 & 82.4 & 96.5 \\
J044354.22$+$125736.7 & L9pec (red) & 49$\pm$5 & 41.5 & 41.4 &  97.8 & 100.6 & 93.4 & 94.7 \\
J045800.69$+$100456.9 & T2pec (red) & 41$\pm$7 & 47.3 & 46.8 &  95.0 &  96.2 & 82.0 & 97.8 \\
J051215.86$+$165818.0 & L9          & 44$\pm$3 & 44.1 & 45.8 & 114.0 & 115.8 & 36.6 & 90.6 \\
\enddata
\tablenotetext{\dagger}{dist$_{\rm phot}$, dist$_{\rm cp}$, and dist$_{\rm BANYAN}$ are the photometric distance estimate, the distance predicted from the convergent point method (assuming Hyades membership), and the BANYAN $\Sigma$ predicted distance (again, assuming Hyades membership), respectively.}
\tablenotetext{\ddagger}{$\theta_{\mu}$ and $\theta_{\rm cp}$ are the proper motion angle and the convergent point angle, as described in the text.}
\tablenotetext{a}{Hyades membership probability from BANYAN $\Sigma$ that does not include a distance estimate as a constraint.}
\tablenotetext{b}{Hyades membership probability fro BANYAN $\Sigma$ that includes the photometric distance.}
\end{deluxetable*}

\subsection{Physical Properties}
\label{sec:properties}

Brown dwarfs do not form a main sequence, but instead cool and dim over time as they radiate the initial heat of their formation \citep{burrows1997}, which leads to a degenerate relationship between mass, luminosity, and age. Thus, precise fundamental properties of most brown dwarfs are exceptionally difficult to determine. However, if a brown dwarf can be tied to a group of stars with a known age like the Hyades, this degeneracy can be broken.  We can therefore estimate the masses of each of our new candidates by using evolutionary models, the age of the Hyades (650$\pm$50 Myr), and effective temperature (\teff) estimates derived from spectral types. 

In this work, \teff\ values are estimated using the spectral type vs.\ \teff\ relation from \cite{kirkpatrick2021}.  All objects are given a $\pm$0.5 spectral subtype uncertainty, with the exceptions of CWISE J041259.89$+$085049.6, CWISE J042222.17$+$213900.5, CWISE J044354.22$+$125736.7, and CWISE J045800.69$+$100456.9, for which we assume spectral type uncertainties of $\pm$1.  \teff\ values are given in Table \ref{tab:phys}. 

\cite{schneider2022} used the evolutionary models of \cite{saumon2008} and \cite{phillips2020} to estimate the masses of newly discovered objects.  In this work, we also use the \cite{phillips2020} models, but use the grid that includes the updated equation of state from \cite{chabrier2023}.  Notably, this updated grid results in slightly higher masses than the previously used \cite{phillips2020} evolutionary models. For example, \cite{schneider2022} found a mass of 33$\pm$3 \mjup\ for the T3 brown dwarf Hyades candidate CWISE J043018.70$+$105857.1 using the \cite{phillips2020} models, where we find 35$\pm$3 \mjup\ using \cite{phillips2020} models with the updated equation of state.  Furthermore, we use the updated Sonora Diamondback evolutionary models of \cite{marley2021} and \cite{morley2024} instead of \cite{saumon2008}.  We use the solar-metallicity hybrid-grav models in this work, which includes gravity-dependent cloud-clearing. Mass estimates assuming Hyades membership are given in Table \ref{tab:phys}.  We note that CWISE J042356.23$+$130414.3 has the latest spectral type of any free-floating Hyades candidate member, and thus the lowest mass estimates of 33$\pm$4 $M_{\rm Jup}$ and 28$^{+7}_{-4}$ $M_{\rm Jup}$ using the \cite{phillips2020} and \cite{morley2024} models, respectively.  

\begin{deluxetable}{lcccccccc}
\label{tab:phys}
\tabletypesize{\footnotesize}
\tablecaption{Physical Properties of New Hyades Candidates}
\tablehead{
\colhead{CWISE} & \colhead{Spec.} & \colhead{\teff} & \colhead{Mass\tablenotemark{a}} & \colhead{Mass\tablenotemark{b}}  \\
\colhead{Name} & \colhead{Type} & \colhead{(K)} & \colhead{($M_{\rm Jup}$)} & \colhead{($M_{\rm Jup}$)}  }
\startdata
J041259.89$+$085049.6 & T3pec (red) & 1202$\pm$81  & 35$\pm$3 & 31$^{+9}_{-4}$ \\
J042222.17$+$213900.5 & L5-L7 (red) & 1515$\pm$165 & 47$\pm$6 & 52$\pm$7 \\
J042356.23$+$130414.3 & T5          & 1136$\pm$115 & 33$\pm$4 & 28$^{+7}_{-4}$ \\
J043511.26$+$213846.3 & L7.5        & 1378$\pm$140 & 42$\pm$5 & 46$^{+6}_{-12}$ \\
J044116.07$+$143645.4 & T2          & 1220$\pm$80  & 35$\pm$3 & 32$^{+10}_{-4}$ \\
J044354.22$+$125736.7 & L9pec (red) & 1275$\pm$81  & 38$\pm$3 & 39$^{+7}_{-8}$ \\
J045800.69$+$100456.9 & T2pec (red) & 1220$\pm$81  & 35$\pm$3 & 32$^{+10}_{-4}$ \\
J051215.86$+$165818.0 & L9          & 1275$\pm$80  & 38$\pm$3 & 39$^{+7}_{-8}$ \\
\enddata
\tablenotetext{a}{Masses determined using the models of \cite{phillips2020} and updated equation of state in \cite{chabrier2023}.}
\tablenotetext{b}{Masses determined using the hybrid-grav models of \cite{morley2024}.}
\end{deluxetable}

\section{Discussion}
\label{sec:disc}

While the age of the Hyades is sufficiently old such that many common spectroscopic indicators of youth (e.g., \citealt{allers2013}) are indistinguishable from field-age objects for Hyades members, we note that the spectra of both Hyades members in \cite{schneider2017} were slightly red compared to spectral standards, while \cite{best2015} found the L6 Hyades member Hya12 to also be unusually red.  Several of the Hyades candidates in this work also appear redder than near-infrared spectral standards (see Figure \ref{fig:spectra}). The distance of the Hyades makes interstellar reddening unlikely to be significant (e.g., \citealt{taylor2006}), and red near-infrared colors are common to young brown dwarfs (e.g., \citealt{faherty2016, liu2016}).

\begin{figure}
\plotone{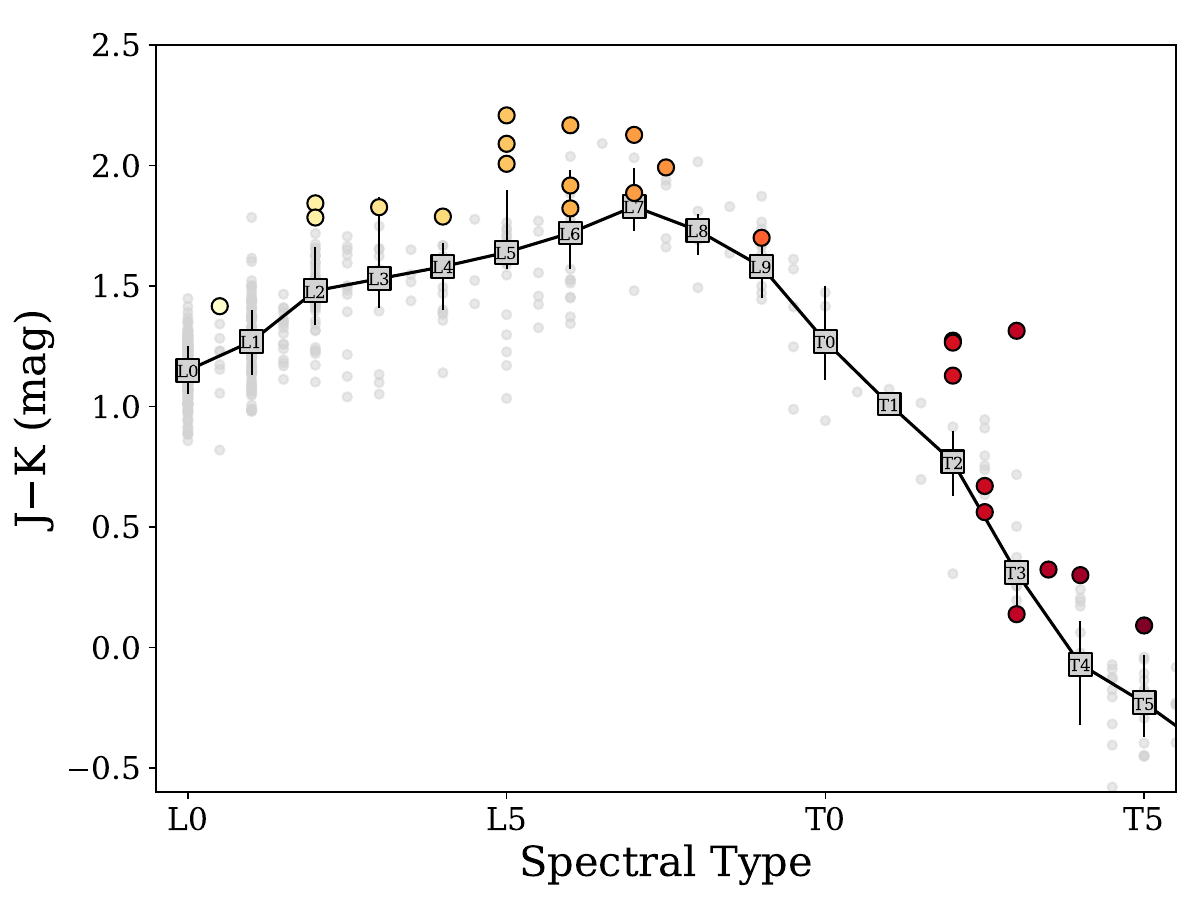}
\caption{Color versus near-infrared spectral type diagram for L and T dwarfs from the UHS survey in \cite{schneider2023} (grey circles) compared to candidate and confirmed Hyades L and T dwarfs (colored circles).  Median colors for each spectral type for field age objects are labeled grey squares. 
The error bars on the median values represent the 16 and 84 percentile ranges as determined in \cite{schneider2023}.}  
\label{fig:colors1}
\end{figure}

With the new discoveries presented in this work, we can attempt to investigate the near-infrared color sequence for Hyades-age brown dwarfs. Figure \ref{fig:colors1} shows a $J-K$ vs.\ spectral type diagram comparing high-probability substellar Hyades members (BANYAN $\Sigma$ $\geq$ 80\%) to field age brown dwarfs from \cite{schneider2023}.  We use the BANYAN $\Sigma$ probabilities from Table 5 of \cite{schneider2022} for previously known Hyades candidates, and the probabilities from Table \ref{tab:member} (without the inclusion of photometric distances) for new candidates from this work.  We chose a high threshold for BANYAN $\Sigma$ probabilities to investigate color trends for the most likely substellar Hyades members and to mitigate the influence of potential interlopers.  There are 25 objects that survive this BANYAN $\Sigma$ probability cut, including PSO J049.1159$+$26.8409 \citep{best2015, zhang2021}, PSO J052.2746$+$13.3754 and PSO J069.7303$+$04.3834 \citep{best2020, zhang2021}, WISEA J041232.77$+$104408.3, WISEA J043642.75$+$190134.8, and WISEA J044105.56$+$213001.5 \citep{schneider2017}, Hya11 \citep{hogan2008, martin2018}, 2MASS J04183483$+$2131275 \citep{perez2017}, 2MASS J04241856+0637448 \citep{perez2018}, CWISE J033817.87$+$171744.1, CWISE J041953.55$+$203628.0, CWISE J042731.38$+$074344.9, and CWISE J043018.70$+$105857.1 \citep{schneider2022}, CFHT-Hy-20 \citep{bouvier2008}, Hya08, Hya09, Hya10, and Hya12 \citep{hogan2008, lodieu2014} and CWISE J041259.89$+$085049.6, CWISE J042222.17$+$213900.5, CWISE J042356.23$+$130414.3, CWISE J043511.26$+$213846.3, CWISE J044116.07$+$143645.4, CWISE J044354.22$+$125736.7, and CWISE J045800.69$+$100456.9 from this work. 

Every candidate L- and T-type Hyades member in this sample, spanning spectral types L0.5 to T5, has a redder $J-K$ color than the median field age-color for that spectral type, with one exception, the T3 dwarf CWISE J043018.70$+$105857.1 \citep{schneider2022}.  The color difference seen for the vast majority of the Hyades candidate sample compared to the field sequence may be due to several factors, including slightly lower surface gravities, enhanced metallicity, viewing angle, and variability.  

Using the \cite{morley2024} hybrid-grav evolutionary models, we find that Hyades-age brown dwarfs have log($g$) values $\sim$0.08 dex and $\sim$0.13 dex lower than 2 Gyr and 5 Gyr objects on average, respectively, for masses between 0.03 and 0.07 $M_{\odot}$.  While minor, this difference in surface gravity could lead to reduced absorption from collisionally-induced H$_2$ (e.g., \citealt{linsky1969}), higher-altitude clouds (e.g., \citealt{madhu2011}), and cloud grain-size or composition differences (e.g., \citealt{suarez2023a}).  

The metallicity for the Hyades has consistently been determined to be slightly metal-rich ([Fe/H] = 0.09--0.24; \citealt{dutra2016, cummings2017, gossage2018, wanderley2023}).  Trends have been shown to exist between metallicity and infrared color excess (e.g., \citealt{looper2008, zhang2024}), with higher-metallicity L-type brown dwarfs having redder colors on average than those brown dwarfs with metallicities closer to solar.  Though the impact of metallicity on the atmospheres of relatively cloud-free T-type brown dwarfs is unclear.  However, the Diamondback models can again help to provide some insight with three different metallicities ($-$0.5, 0.0, $+$0.5) available in that grid.  Using synthetic photometry, we find the average $J-K$ color difference between [M/H] = $+$0.5 and [M/H] = 0.0 models to be 0.12 mag for substellar objects with log(g) = 5.0, $f_{\rm sed}$ = 3, and \teff\ values between 1200 K and 2200K.  Therefore, the Diamondback models support the empirical metallicity versus color trends, with higher metallicity objects having redder near-infrared colors.      

Further, whether a brown dwarf is viewed closer to edge-on or pole-on can also affect near-infrared colors (e.g., \citealt{vos2017, toro2022, suarez2023b}).  While spin-alignment for stellar members has been found in some clusters (e.g., \citealt{corsaro2017}, other inclination studies of open clusters have found that the alignment of rotation axes is not ubiquitous (e.g., \citealt{jackson2010, healy2023}).  Such a study of the stellar and substellar population of the Hyades would help to determine the influence of viewing angle on the observed $J-K$ color offset for substellar Hyades members.  

Lastly, variability (e.g., \citealt{ashraf2022, oliveros2022}) and unresolved binarity (e.g., \citealt{burgasser2010, dupuy2012}) can also affect near-infrared colors, which we showed is the potential reason for the unusually red spectra of CWISE J041259.89$+$085049.6 and CWISE J045800.69$+$100456.9 in Section \ref{sec:spts}.  Brown dwarf binary fractions are 10--20\% for field age objects \citep{burgasser2006b, radigan2013, aberasturi2014, fontanive2018, fontanive2023}, but are much higher at young ages \citep{defurio2022}.  The binary fraction of Hyades-age brown dwarfs is currently unknown, but unresolved binarity could also contribute to the observed $J-K$ color excess. We note that two of the high-probability Hyades members from \cite{zhang2021} (PSO J049.1159$+$26.8409 and PSO J069.7303$+$04.3834) were found to be strong binary candidates, while CFHT-Hy-20 \citep{bouvier2008} was found to be a weak binary candidate in that work.

Additional observations of these candidates, including high-resolution imaging, variability monitoring, and/or James Webb Space Telescope spectra of silicate features in the mid-infrared (e.g., \citealt{miles2023}) could help to shed light on the impact of binarity and cloud properties of these intermediate-age brown dwarfs.

\section{Summary}
\label{sec:summary}

We have presented eight new candidate substellar Hyades members with spectral types ranging from L6 to T5.  The positions, proper motions, and distance estimates of all eight candidates are supportive of Hyades membership, with parallax and radial velocity measurements needed for full confirmation. If confirmed, CWISE J042356.23$+$130414.3 would be the latest spectral type, and thus lowest-mass free-floating member of the Hyades yet discovered.  We find that two objects (CWISE J041259.89$+$085049.6 and CWISE J045800.69$+$100456.9) have spectra suggestive of near-infrared variability.  An investigation of the near-infrared colors of the highest-probability Hyades candidates reveals that their near-infrared $J-K$ colors are almost always redder than field-age objects of the same spectral type, potentially due to lower surface gravities and enhanced metallicity.

The primary limitation of this survey was the requirement that $J$-band magnitudes be brighter than 18 mag.  Using the absolute $J$-band versus spectral type relations for UHS photometry from \cite{schneider2023} shows that a magnitude limit of 18 can detect L5, T0, and T5 type brown dwarfs out to $\sim$80 pc, $\sim$52 pc, and $\sim$49 pc, respectively.  Note, however that the red colors of Hyades members discussed in Section \ref{sec:disc} suggests that these distance limits are potentially upper limits. Assuming the Hyades is roughly spherical with a tidal radius of 18 pc and an average distance of 47.9 pc \citep{gaia2021}, the 18 mag distance limits suggest that we could be complete at spectral type L5, but are only reaching out to the approximate middle of the cluster distance for T dwarfs. The 5$\sigma$ point source sensitivity of UHS survey $J$-band observations is 19.6 mag \citep{dye2018}.  Extending this type of search to the UHS $J$-band limit results in distance limits of $\sim$159 pc for L5 dwarfs, $\sim$103 pc for T0 dwarfs, and $\sim$97 pc for T5 dwarfs.  However, while we did not enforce a strict $K$-band limit, surveys for new later-type brown dwarfs in the Hyades may be limited by $K$-band photometry rather than $J$, as brown dwarfs beyond the L/T transition trend toward bluer $J-K$ colors.  Using a 5$\sigma$ point source sensitivity $K$-band limit of 18.4 mag (Bruursema et al. in prep.), we find limits of $\sim$204 pc, $\sim$103 pc, and $\sim$53 pc for L5, T0, and T5 dwarfs, respectively. Regardless, pushing this type of survey to deeper limits available with UHS would help to create a more complete census of brown dwarfs in the Hyades, especially at later spectral types ($\geq$T0).   

\acknowledgments

The authors wish to thank Caroline Morley for making available the Sonora Diamondback models.  This publication makes use of data products from the UKIRT Hemisphere Survey, which is a joint project of the United States Naval Observatory, the University of Hawaii Institute for Astronomy, the Cambridge University Cambridge Astronomy Survey Unit, and the University of Edinburgh Wide-Field Astronomy Unit (WFAU). This project was primarily funded by the United States Navy. The WFAU gratefully acknowledges support for this work from the Science and Technology Facilities Council through ST/T002956/1 and previous grants. 

This work is based on observations obtained at the international Gemini Observatory, a program of NSF’s NOIRLab, which is managed by the Association of Universities for Research in Astronomy (AURA) under a cooperative agreement with the National Science Foundation on behalf of the Gemini Observatory partnership: the National Science Foundation (United States), National Research Council (Canada), Agencia Nacional de Investigaci\'{o}n y Desarrollo (Chile), Ministerio de Ciencia, Tecnolog\'{i}a e Innovaci\'{o}n (Argentina), Minist\'{e}rio da Ci\^{e}ncia, Tecnologia, Inova\c{c}\~{o}es e Comunica\c{c}\~{o}es (Brazil), and Korea Astronomy and Space Science Institute (Republic of Korea).  

This publication makes use of data products from the {\it Wide-field Infrared Survey Explorer}, which is a joint project of the University of California, Los Angeles, and the Jet Propulsion Laboratory/California Institute of Technology, and NEOWISE which is a project of the Jet Propulsion Laboratory/California Institute of Technology. {\it WISE} and NEOWISE are funded by the National Aeronautics and Space Administration.  Part of this research was carried out at the Jet Propulsion Laboratory, California Institute of Technology, under a contract with the National Aeronautics and Space Administration.  

The Pan-STARRS1 Surveys (PS1) and the PS1 public science archive have been made possible through contributions by the Institute for Astronomy, the University of Hawaii, the Pan-STARRS Project Office, the Max-Planck Society and its participating institutes, the Max Planck Institute for Astronomy, Heidelberg and the Max Planck Institute for Extraterrestrial Physics, Garching, The Johns Hopkins University, Durham University, the University of Edinburgh, the Queen's University Belfast, the Harvard-Smithsonian Center for Astrophysics, the Las Cumbres Observatory Global Telescope Network Incorporated, the National Central University of Taiwan, the Space Telescope Science Institute, the National Aeronautics and Space Administration under Grant No. NNX08AR22G issued through the Planetary Science Division of the NASA Science Mission Directorate, the National Science Foundation Grant No. AST-1238877, the University of Maryland, Eotvos Lorand University (ELTE), the Los Alamos National Laboratory, and the Gordon and Betty Moore Foundation.

This work has benefitted from The UltracoolSheet at \href{http://bit.ly/UltracoolSheet}{http://bit.ly/UltracoolSheet}, maintained by Will Best, Trent Dupuy, Michael Liu, Aniket Sanghi, Rob Siverd, and Zhoujian Zhang, and developed from compilations by \cite{dupuy2012} \cite{dupuy2013}, \cite{deacon2014}, \cite{liu2016}, \cite{best2018}, \cite{best2021}, \cite{schneider2023}, and \cite{sanghi2023}.

The authors wish to recognize and acknowledge the very significant cultural role and reverence that the summit of Maunakea has always had within the indigenous Hawaiian community.  We are most fortunate to have the opportunity to conduct observations from this mountain.

\software{BANYAN~$\Sigma$ \citep{gagne2018}, pypeit \citep{prochaska2020, prochaska2023}}



\begin{thebibliography}{}
\bibitem[Aberasturi et al.(2014)]{aberasturi2014} Aberasturi, M., Burgasser, A.~J., Mora, A., et al.\ 2014, \aj, 148, 129. doi:10.1088/0004-6256/148/6/129
\bibitem[Al-Johani et al.(2022)]{aljohani2022} Al-Johani, A.~S., Elsanhoury, W.~H., Al-Juhani, A., et al.\ 2022, Kinematics and Physics of Celestial Bodies, 38, 240. doi:10.3103/S0884591322050026
\bibitem[Allers \& Liu(2013)]{allers2013} Allers, K.~N. \& Liu, M.~C.\ 2013, \apj, 772, 79. doi:10.1088/0004-637X/772/2/79
\bibitem[Ashraf et al.(2022)]{ashraf2022} Ashraf, A., Bardalez Gagliuffi, D.~C., Manjavacas, E., et al.\ 2022, \apj, 934, 178. doi:10.3847/1538-4357/ac7aab
\bibitem[Bannister \& Jameson(2007)]{bannister2007} Bannister, N.~P. \& Jameson, R.~F.\ 2007, \mnras, 378, L24. doi:10.1111/j.1745-3933.2007.00312.x
\bibitem[Bardalez Gagliuffi et al.(2014)]{bardalez2014} Bardalez Gagliuffi, D.~C., Burgasser, A.~J., Gelino, C.~R., et al.\ 2014, \apj, 794, 143. doi:10.1088/0004-637X/794/2/143
\bibitem[Bell et al.(2015)]{bell2015} Bell, C.~P.~M., Mamajek, E.~E., \& Naylor, T.\ 2015, \mnras, 454, 593. doi:10.1093/mnras/stv1981
\bibitem[Best et al.(2015)]{best2015} Best, W.~M.~J., Liu, M.~C., Magnier, E.~A., et al.\ 2015, \apj, 814, 118. doi:10.1088/0004-637X/814/2/118
\bibitem[Best et al.(2018)]{best2018} Best, W.~M.~J., Magnier, E.~A., Liu, M.~C., et al.\ 2018, \apjs, 234, 1. doi:10.3847/1538-4365/aa9982
\bibitem[Best et al.(2020)]{best2020} Best, W.~M.~J., Liu, M.~C., Magnier, E.~A., et al.\ 2020, \aj, 159, 257. doi:10.3847/1538-3881/ab84f4
\bibitem[Best et al.(2021)]{best2021} Best, W.~M.~J., Liu, M.~C., Magnier, E.~A., et al.\ 2021, \aj, 161, 42. doi:10.3847/1538-3881/abc893
\bibitem[Boss(1908)]{boss1908} Boss, L.~J.\ 1908, \aj, 26, 31. doi:10.1086/103802
\bibitem[Bouvier et al.(2008)]{bouvier2008} Bouvier, J., Kendall, T., Meeus, G., et al.\ 2008, \aap, 481, 661. doi:10.1051/0004-6361:20079303
\bibitem[Brandt \& Huang(2015a)]{brandt2015a} Brandt, T.~D. \& Huang, C.~X.\ 2015a, \apj, 807, 24. doi:10.1088/0004-637X/807/1/24
\bibitem[Brandt \& Huang(2015b)]{brandt2015b} Brandt, T.~D. \& Huang, C.~X.\ 2015b, \apj, 807, 58. doi:10.1088/0004-637X/807/1/58
\bibitem[Bravo et al.(2023)]{bravo2023} Bravo, A., Schneider, A.~C., Bardalez Gagliuffi, D., et al.\ 2023, \aj, 166, 226. doi:10.3847/1538-3881/acffc1
\bibitem[Bryja et al.(1992)]{bryja1992} Bryja, C., Jones, T.~J., Humphreys, R.~M., et al.\ 1992, \apjl, 388, L23. doi:10.1086/186321
\bibitem[Bryja et al.(1994)]{bryja1994} Bryja, C., Humphreys, R.~M., \& Jones, T.~J.\ 1994, \aj, 107, 246. doi:10.1086/116848
\bibitem[Burgasser et al.(2002)]{burgasser2002} Burgasser, A.~J., Kirkpatrick, J.~D., Brown, M.~E., et al.\ 2002, \apj, 564, 421. doi:10.1086/324033
\bibitem[Burgasser et al.(2004)]{burgasser2004} Burgasser, A.~J., McElwain, M.~W., Kirkpatrick, J.~D., et al.\ 2004, \aj, 127, 2856. doi:10.1086/383549
\bibitem[Burgasser et al.(2006a)]{burgasser2006} Burgasser, A.~J., Geballe, T.~R., Leggett, S.~K., et al.\ 2006a, \apj, 637, 1067. doi:10.1086/498563
\bibitem[Burgasser et al.(2006b)]{burgasser2006b} Burgasser, A.~J., Kirkpatrick, J.~D., Cruz, K.~L., et al.\ 2006b, \apjs, 166, 585. doi:10.1086/506327
\bibitem[Burgasser(2007)]{burgasser2007} Burgasser, A.~J.\ 2007, \apj, 659, 655. doi:10.1086/511027
\bibitem[Burgasser et al.(2010)]{burgasser2010} Burgasser, A.~J., Cruz, K.~L., Cushing, M., et al.\ 2010, \apj, 710, 1142. doi:10.1088/0004-637X/710/2/1142
\bibitem[Burrows et al.(1997)]{burrows1997} Burrows, A., Marley, M., Hubbard, W.~B., et al.\ 1997, \apj, 491, 856. doi:10.1086/305002
\bibitem[Canty et al.(2013)]{canty2013} Canty, J.~I., Lucas, P.~W., Roche, P.~F., et al.\ 2013, \mnras, 435, 2650. doi:10.1093/mnras/stt1477
\bibitem[Casewell et al.(2014)]{casewell2014} Casewell, S.~L., Littlefair, S.~P., Burleigh, M.~R., et al.\ 2014, \mnras, 441, 2644. doi:10.1093/mnras/stu746
\bibitem[Chambers et al.(2016)]{chambers2016} Chambers, K.~C., Magnier, E.~A., Metcalfe, N., et al.\ 2016, arXiv:1612.05560. doi:10.48550/arXiv.1612.05560
\bibitem[Chabrier et al.(2023)]{chabrier2023} Chabrier, G., Baraffe, I., Phillips, M., et al.\ 2023, \aap, 671, A119. doi:10.1051/0004-6361/202243832
\bibitem[Corsaro et al.(2017)]{corsaro2017} Corsaro, E., Lee, Y.-N., Garc{\'\i}a, R.~A., et al.\ 2017, Nature Astronomy, 1, 0064. doi:10.1038/s41550-017-0064
\bibitem[Cummings et al.(2017)]{cummings2017} Cummings, J.~D., Deliyannis, C.~P., Maderak, R.~M., et al.\ 2017, \aj, 153, 128. doi:10.3847/1538-3881/aa5b86
\bibitem[Deacon et al.(2014)]{deacon2014} Deacon, N.~R., Liu, M.~C., Magnier, E.~A., et al.\ 2014, \apj, 792, 119. doi:10.1088/0004-637X/792/2/119
\bibitem[de Bruijne et al.(2001)]{debruijne2001} de Bruijne, J.~H.~J., Hoogerwerf, R., \& de Zeeuw, P.~T.\ 2001, \aap, 367, 111. doi:10.1051/0004-6361:20000410
\bibitem[De Furio et al.(2022)]{defurio2022} De Furio, M., Liu, C., Meyer, M.~R., et al.\ 2022, \apj, 941, 161. doi:10.3847/1538-4357/aca285
\bibitem[De Gennaro et al.(2009)]{degennaro2009} De Gennaro, S., von Hippel, T., Jefferys, W.~H., et al.\ 2009, \apj, 696, 12. doi:10.1088/0004-637X/696/1/12
\bibitem[Dobbie et al.(2002)]{dobbie2002} Dobbie, P.~D., Kenyon, F., Jameson, R.~F., et al.\ 2002, \mnras, 329, 543. doi:10.1046/j.1365-8711.2002.05002.x
\bibitem[Dupuy \& Liu(2012)]{dupuy2012} Dupuy, T.~J. \& Liu, M.~C.\ 2012, \apjs, 201, 19. doi:10.1088/0067-0049/201/2/19
\bibitem[Dupuy \& Kraus(2013)]{dupuy2013} Dupuy, T.~J. \& Kraus, A.~L.\ 2013, Science, 341, 1492. doi:10.1126/science.1241917
\bibitem[Dutra-Ferreira et al.(2016)]{dutra2016} Dutra-Ferreira, L., Pasquini, L., Smiljanic, R., et al.\ 2016, \aap, 585, A75. doi:10.1051/0004-6361/201526783
\bibitem[Dye et al.(2018)]{dye2018} Dye, S., Lawrence, A., Read, M.~A., et al.\ 2018, \mnras, 473, 5113. doi:10.1093/mnras/stx2622
\bibitem[Elias et al.(2006)]{elias2006} Elias, J.~H., Joyce, R.~R., Liang, M., et al.\ 2006, \procspie, 6269, 62694C. doi:10.1117/12.671817
\bibitem[Faherty et al.(2016)]{faherty2016} Faherty, J.~K., Riedel, A.~R., Cruz, K.~L., et al.\ 2016, \apjs, 225, 10. doi:10.3847/0067-0049/225/1/10
\bibitem[Fontanive et al.(2018)]{fontanive2018} Fontanive, C., Biller, B., Bonavita, M., et al.\ 2018, \mnras, 479, 2702. doi:10.1093/mnras/sty1682
\bibitem[Fontanive et al.(2023)]{fontanive2023} Fontanive, C., Bedin, L.~R., De Furio, M., et al.\ 2023, \mnras, 526, 1783. doi:10.1093/mnras/stad2870
\bibitem[Franson et al.(2023)]{franson2023} Franson, K., Bowler, B.~P., Bonavita, M., et al.\ 2023, \aj, 165, 39. doi:10.3847/1538-3881/aca408
\bibitem[Gagn{\'e} et al.(2018)]{gagne2018} Gagn{\'e}, J., Mamajek, E.~E., Malo, L., et al.\ 2018, \apj, 856, 23. doi:10.3847/1538-4357/aaae09
\bibitem[Gaia Collaboration et al.(2021)]{gaia2021} Gaia Collaboration, Smart, R.~L., Sarro, L.~M., et al.\ 2021, \aap, 649, A6. doi:10.1051/0004-6361/202039498
\bibitem[Gaia Collaboration et al.(2023)]{gaia2023} Gaia Collaboration, Vallenari, A., Brown, A.~G.~A., et al.\ 2023, \aap, 674, A1. doi:10.1051/0004-6361/202243940
\bibitem[Giclas et al.(1962)]{giclas1962} Giclas, H.~L., Burnham, R., \& Thomas, N.~G.\ 1962, Lowell Observatory Bulletin, 5, 257
\bibitem[Gizis et al.(1999)]{gizis1999} Gizis, J.~E., Reid, I.~N., \& Monet, D.~G.\ 1999, \aj, 118, 997. doi:10.1086/300982
\bibitem[Goldman et al.(2013)]{goldman2013} Goldman, B., R{\"o}ser, S., Schilbach, E., et al.\ 2013, \aap, 559, A43. doi:10.1051/0004-6361/201321727
\bibitem[Gossage et al.(2018)]{gossage2018} Gossage, S., Conroy, C., Dotter, A., et al.\ 2018, \apj, 863, 67. doi:10.3847/1538-4357/aad0a0
\bibitem[Guerra Toro et al.(2022)]{toro2022} Guerra Toro, M.~E., Zhou, Y., \& Bowler, B.~P.\ 2022, Research Notes of the American Astronomical Society, 6, 250. doi:10.3847/2515-5172/aca661
\bibitem[Hanson(1975)]{hanson1975} Hanson, R.~B.\ 1975, \aj, 80, 379. doi:10.1086/111753
\bibitem[Harris et al.(1999)]{harris1999} Harris, H.~C., Vrba, F.~J., Dahn, C.~C., et al.\ 1999, \aj, 117, 339. doi:10.1086/300692
\bibitem[Hayashi \& Nakano(1963)]{hayashi1963} Hayashi, C. \& Nakano, T.\ 1963, Progress of Theoretical Physics, 30, 460. doi:10.1143/PTP.30.460
\bibitem[Healy et al.(2023)]{healy2023} Healy, B.~F., McCullough, P.~R., Schlaufman, K.~C., et al.\ 2023, \apj, 944, 39. doi:10.3847/1538-4357/acad7b
\bibitem[Herbig(1962)]{herbig1962} Herbig, G.~H.\ 1962, \apj, 135, 736. doi:10.1086/147316
\bibitem[Hertzsprung(1921)]{hertzsprung1921} Hertzsprung, E.\ 1921, \bain, 1, 4
\bibitem[Hodierna(1654)]{hodierna1654} Hodierna, G.~B.\ 1654, De systemate orbis cometici deque admirandis coeli characteribus, opuscula duo, in quorum primo cometarum causae disquiruntur, \& explicantur, necnon vie Com etarum, per orbem cometicum multiplices indicantur. In secundo vero quid, quales, quotue sint stellae luminosae, nebulae, necnon, \& occultae, manifestantur \& rerum caelestium studiosis commendantur, by Hodierna, Giovanni Battista, 1654.. doi:10.3931/e-rara-444
\bibitem[Hogan et al.(2008)]{hogan2008} Hogan, E., Jameson, R.~F., Casewell, S.~L., et al.\ 2008, \mnras, 388, 495. doi:10.1111/j.1365-2966.2008.13437.x
\bibitem[Irwin et al.(2004)]{irwin2004} Irwin, M.~J., Lewis, J., Hodgkin, S., et al.\ 2004, \procspie, 5493, 411. doi:10.1117/12.551449
\bibitem[Jackson \& Jeffries(2010)]{jackson2010} Jackson, R.~J. \& Jeffries, R.~D.\ 2010, \mnras, 402, 1380. doi:10.1111/j.1365-2966.2009.15983.x
\bibitem[Jerabkova et al.(2021)]{jerabkova2021} Jerabkova, T., Boffin, H.~M.~J., Beccari, G., et al.\ 2021, \aap, 647, A137. doi:10.1051/0004-6361/202039949
\bibitem[Johnson \& Knuckles(1955)]{johnson1955} Johnson, H.~L. \& Knuckles, C.~F.\ 1955, \apj, 122, 209. doi:10.1086/146079
\bibitem[Kapteyn \& Desetter(1904)]{kapteyn1904} Kapteyn, J.~C. \& Desetter, W.\ 1904, Publications of the Kapteyn Astronomical Laboratory Groningen, 14, I
\bibitem[Kirkpatrick et al.(1999)]{kirkpatrick1999} Kirkpatrick, J.~D., Reid, I.~N., Liebert, J., et al.\ 1999, \apj, 519, 802. doi:10.1086/307414
\bibitem[Kirkpatrick et al.(2000)]{kirkpatrick2000} Kirkpatrick, J.~D., Reid, I.~N., Liebert, J., et al.\ 2000, \aj, 120, 447. doi:10.1086/301427
\bibitem[Kirkpatrick et al.(2010)]{kirkpatrick2010} Kirkpatrick, J.~D., Looper, D.~L., Burgasser, A.~J., et al.\ 2010, \apjs, 190, 100. doi:10.1088/0067-0049/190/1/100
\bibitem[Kirkpatrick et al.(2011)]{kirkpatrick2011} Kirkpatrick, J.~D., Cushing, M.~C., Gelino, C.~R., et al.\ 2011, \apjs, 197, 19. doi:10.1088/0067-0049/197/2/19
\bibitem[Kirkpatrick et al.(2021)]{kirkpatrick2021} Kirkpatrick, J.~D., Gelino, C.~R., Faherty, J.~K., et al.\ 2021, \apjs, 253, 7. doi:10.3847/1538-4365/abd107
\bibitem[Kuchner et al.(2017)]{kuchner2017} Kuchner, M.~J., Faherty, J.~K., Schneider, A.~C., et al.\ 2017, \apjl, 841, L19. doi:10.3847/2041-8213/aa7200
\bibitem[Kumar(1963)]{kumar1963} Kumar, S.~S.\ 1963, \apj, 137, 1121. doi:10.1086/147589
\bibitem[K{\"u}stner(1902)]{kustner1902} K{\"u}stner, F.\ 1902, Astronomische Nachrichten, 158, 359. doi:10.1002/asna.19021582303
\bibitem[Kuzuhara et al.(2022)]{kuzuhara2022} Kuzuhara, M., Currie, T., Takarada, T., et al.\ 2022, \apjl, 934, L18. doi:10.3847/2041-8213/ac772f
\bibitem[Lawrence et al.(2007)]{lawrence2007} Lawrence, A., Warren, S.~J., Almaini, O., et al.\ 2007, \mnras, 379, 1599. doi:10.1111/j.1365-2966.2007.12040.x
\bibitem[Lebreton et al.(2001)]{lebreton2001} Lebreton, Y., Fernandes, J., \& Lejeune, T.\ 2001, \aap, 374, 540. doi:10.1051/0004-6361:20010757
\bibitem[Leggett \& Hawkins(1989)]{leggett1989} Leggett, S.~K. \& Hawkins, M.~R.~S.\ 1989, \mnras, 238, 145. doi:10.1093/mnras/238.1.145
\bibitem[Leggett et al.(1994)]{leggett1994} Leggett, S.~K., Harris, H.~C., \& Dahn, C.~C.\ 1994, \aj, 108, 944. doi:10.1086/117124
\bibitem[Leggett et al.(2000)]{leggett2000} Leggett, S.~K., Geballe, T.~R., Fan, X., et al.\ 2000, \apjl, 536, L35. doi:10.1086/312728
\bibitem[Linsky(1969)]{linsky1969} Linsky, J.~L.\ 1969, \apj, 156, 989. doi:10.1086/150030
\bibitem[Liu et al.(2016)]{liu2016} Liu, M.~C., Dupuy, T.~J., \& Allers, K.~N.\ 2016, \apj, 833, 96. doi:10.3847/1538-4357/833/1/96
\bibitem[Lodieu et al.(2014)]{lodieu2014} Lodieu, N., Boudreault, S., \& B{\'e}jar, V.~J.~S.\ 2014, \mnras, 445, 3908. doi:10.1093/mnras/stu2059
\bibitem[Lodieu et al.(2019)]{lodieu2019} Lodieu, N., Smart, R.~L., P{\'e}rez-Garrido, A., et al.\ 2019, \aap, 623, A35. doi:10.1051/0004-6361/201834045
\bibitem[Lodieu(2020)]{lodieu2020} Lodieu, N.\ 2020, \memsai, 91, 84
\bibitem[Looper et al.(2008)]{looper2008} Looper, D.~L., Kirkpatrick, J.~D., Cutri, R.~M., et al.\ 2008, \apj, 686, 528. doi:10.1086/591025
\bibitem[Lucas et al.(2008)]{lucas2008} Lucas, P.~W., Hoare, M.~G., Longmore, A., et al.\ 2008, \mnras, 391, 136. doi:10.1111/j.1365-2966.2008.13924.x
\bibitem[Luyten \& Merrill(1954)]{luyten1954} Luyten, W.~J. \& Merrill, P.~W.\ 1954, \pasp, 66, 207. doi:10.1086/126697
\bibitem[Madhusudhan et al.(2011)]{madhu2011} Madhusudhan, N., Burrows, A., \& Currie, T.\ 2011, \apj, 737, 34. doi:10.1088/0004-637X/737/1/34
\bibitem[Madsen et al.(2002)]{madsen2002} Madsen, S., Dravins, D., \& Lindegren, L.\ 2002, \aap, 381, 446. doi:10.1051/0004-6361:20011458
\bibitem[Magnier et al.(2020)]{magnier2020} Magnier, E.~A., Schlafly, E.~F., Finkbeiner, D.~P., et al.\ 2020, \apjs, 251, 6. doi:10.3847/1538-4365/abb82a
\bibitem[Marley et al.(2021)]{marley2021} Marley, M.~S., Saumon, D., Visscher, C., et al.\ 2021, \apj, 920, 85. doi:10.3847/1538-4357/ac141d
\bibitem[Marocco et al.(2014)]{marocco2014} Marocco, F., Day-Jones, A.~C., Lucas, P.~W., et al.\ 2014, \mnras, 439, 372. doi:10.1093/mnras/stt2463
\bibitem[Marocco et al.(2021)]{marocco2021} Marocco, F., Eisenhardt, P.~R.~M., Fowler, J.~W., et al.\ 2021, \apjs, 253, 8. doi:10.3847/1538-4365/abd805
\bibitem[Mart{\'\i}n et al.(2018)]{martin2018} Mart{\'\i}n, E.~L., Lodieu, N., Pavlenko, Y., et al.\ 2018, \apj, 856, 40. doi:10.3847/1538-4357/aaaeb8
\bibitem[Meingast \& Alves(2019)]{meingast2019} Meingast, S. \& Alves, J.\ 2019, \aap, 621, L3. doi:10.1051/0004-6361/201834622
\bibitem[Melnikov \& Eisl{\"o}ffel(2018)]{melnikov2018} Melnikov, S. \& Eisl{\"o}ffel, J.\ 2018, \aap, 611, A34. doi:10.1051/0004-6361/201630134
\bibitem[Miles et al.(2023)]{miles2023} Miles, B.~E., Biller, B.~A., Patapis, P., et al.\ 2023, \apjl, 946, L6. doi:10.3847/2041-8213/acb04a
\bibitem[Morley et al.(2024)]{morley2024} Morley, C.~V., Mukherjee, S., Marley, M.~S., et al.\ 2024, arXiv:2402.00758. doi:10.48550/arXiv.2402.00758
\bibitem[Oliveros-Gomez et al.(2022)]{oliveros2022} Oliveros-Gomez, N., Manjavacas, E., Ashraf, A., et al.\ 2022, \apj, 939, 72. doi:10.3847/1538-4357/ac96f2
\bibitem[Oort(1979)]{oort1979} Oort, J.~H.\ 1979, \aap, 78, 312
\bibitem[Pels et al.(1975)]{pels1975} Pels, G., Oort, J.~H., \& Pels-Kluyver, H.~A.\ 1975, \aap, 43, 423
\bibitem[P{\'e}rez-Garrido et al.(2017)]{perez2017} P{\'e}rez-Garrido, A., Lodieu, N., \& Rebolo, R.\ 2017, \aap, 599, A78. doi:10.1051/0004-6361/201628778
\bibitem[P{\'e}rez-Garrido et al.(2018)]{perez2018} P{\'e}rez-Garrido, A., Lodieu, N., Rebolo, R., et al.\ 2018, \aap, 620, A130. doi:10.1051/0004-6361/201833672
\bibitem[Perryman et al.(1998)]{perryman1998} Perryman, M.~A.~C., Brown, A.~G.~A., Lebreton, Y., et al.\ 1998, \aap, 331, 81. doi:10.48550/arXiv.astro-ph/9707253
\bibitem[Pesch(1968)]{pesch1968} Pesch, P.\ 1968, \apj, 151, 605. doi:10.1086/149460
\bibitem[Phillips et al.(2020)]{phillips2020} Phillips, M.~W., Tremblin, P., Baraffe, I., et al.\ 2020, \aap, 637, A38. doi:10.1051/0004-6361/201937381
\bibitem[Prochaska et al.(2020)]{prochaska2020} Prochaska, J., Hennawi, J., Westfall, K., et al.\ 2020, The Journal of Open Source Software, 5, 2308. doi:10.21105/joss.02308
\bibitem[Prochaska et al.(2023)]{prochaska2023} Prochaska, J.~X., Hennawi, J., Cooke, R., et al.\ 2023, Zenodo
\bibitem[Proctor(1870)]{proctor1870} Proctor, R.~A.\ 1870, Proceedings of the Royal Society of London Series I, 18, 169. doi:10.1098/rspl.1869.0039
\bibitem[Radigan et al.(2012)]{radigan2012} Radigan, J., Jayawardhana, R., Lafreni{\`e}re, D., et al.\ 2012, \apj, 750, 105. doi:10.1088/0004-637X/750/2/105
\bibitem[Radigan et al.(2013)]{radigan2013} Radigan, J., Jayawardhana, R., Lafreni{\`e}re, D., et al.\ 2013, \apj, 778, 36. doi:10.1088/0004-637X/778/1/36
\bibitem[Ramberg(1941)]{ramberg1941} Ramberg, J.~M.\ 1941, Stockholms Observatoriums Annaler, 13, 9.1
\bibitem[Reid(1992)]{reid1992} Reid, N.\ 1992, \mnras, 257, 257. doi:10.1093/mnras/257.2.257
\bibitem[Reid(1993)]{reid1993} Reid, N.\ 1993, \mnras, 265, 785. doi:10.1093/mnras/265.4.785
\bibitem[Reid \& Gizis(1997)]{reid1997} Reid, I.~N. \& Gizis, J.~E.\ 1997, \aj, 114, 1992. doi:10.1086/118620
\bibitem[Reid \& Hawley(1999)]{reid1999} Reid, I.~N. \& Hawley, S.~L.\ 1999, \aj, 117, 343. doi:10.1086/300686
\bibitem[Reid \& Mahoney(2000)]{reid2000} Reid, I.~N. \& Mahoney, S.\ 2000, \mnras, 316, 827. doi:10.1046/j.1365-8711.2000.03598.x
\bibitem[Reid et al.(2006)]{reid2006} Reid, I.~N., Lewitus, E., Burgasser, A.~J., et al.\ 2006, \apj, 639, 1114. doi:10.1086/499484
\bibitem[Reino et al.(2018)]{reino2018} Reino, S., de Bruijne, J., Zari, E., et al.\ 2018, \mnras, 477, 3197. doi:10.1093/mnras/sty793
\bibitem[R{\"o}ser et al.(2011)]{roeser2011} R{\"o}ser, S., Schilbach, E., Piskunov, A.~E., et al.\ 2011, \aap, 531, A92. doi:10.1051/0004-6361/201116948
\bibitem[R{\"o}ser et al.(2019)]{roeser2019} R{\"o}ser, S., Schilbach, E., \& Goldman, B.\ 2019, \aap, 621, L2. doi:10.1051/0004-6361/201834608
\bibitem[Sanghi et al.(2023)]{sanghi2023} Sanghi, A., Liu, M.~C., Best, W.~M., et al.\ 2023, arXiv:2309.03082. doi:10.48550/arXiv.2309.03082
\bibitem[Saumon \& Marley(2008)]{saumon2008} Saumon, D. \& Marley, M.~S.\ 2008, \apj, 689, 1327. doi:10.1086/592734
\bibitem[Schneider et al.(2014)]{schneider2014} Schneider, A.~C., Cushing, M.~C., Kirkpatrick, J.~D., et al.\ 2014, \aj, 147, 34. doi:10.1088/0004-6256/147/2/34
\bibitem[Schneider et al.(2017)]{schneider2017} Schneider, A.~C., Windsor, J., Cushing, M.~C., et al.\ 2017, \aj, 153, 196. doi:10.3847/1538-3881/aa6624
\bibitem[Schneider et al.(2022)]{schneider2022} Schneider, A.~C., Vrba, F.~J., Munn, J.~A., et al.\ 2022, \aj, 163, 242. doi:10.3847/1538-3881/ac5f50
\bibitem[Schneider et al.(2023)]{schneider2023} Schneider, A.~C., Munn, J.~A., Vrba, F.~J., et al.\ 2023, \aj, 166, 103. doi:10.3847/1538-3881/ace9bf
\bibitem[Skrutskie et al.(2006)]{skrutskie2006} Skrutskie, M.~F., Cutri, R.~M., Stiening, R., et al.\ 2006, \aj, 131, 1163. doi:10.1086/498708
\bibitem[Su{\'a}rez \& Metchev(2023)]{suarez2023a} Su{\'a}rez, G. \& Metchev, S.\ 2023, \mnras, 523, 4739. doi:10.1093/mnras/stad1711
\bibitem[Su{\'a}rez et al.(2023)]{suarez2023b} Su{\'a}rez, G., Vos, J.~M., Metchev, S., et al.\ 2023, \apjl, 954, L6. doi:10.3847/2041-8213/acec4b
\bibitem[Stauffer(1982)]{stauffer1982} Stauffer, J.\ 1982, \aj, 87, 899. doi:10.1086/113171
\bibitem[Stauffer et al.(1994)]{stauffer1994} Stauffer, J.~R., Liebert, J., Giampapa, M., et al.\ 1994, \aj, 108, 160. doi:10.1086/117054
\bibitem[Stauffer et al.(1995)]{stauffer1995} Stauffer, J.~R., Liebert, J., \& Giampapa, M.\ 1995, \aj, 109, 298. doi:10.1086/117273
\bibitem[Taylor(2006)]{taylor2006} Taylor, B.~J.\ 2006, \aj, 132, 2453. doi:10.1086/508610
\bibitem[Titus \& Morgan(1940)]{titus1940} Titus, J. \& Morgan, W.~W.\ 1940, \apj, 92, 256. doi:10.1086/144215
\bibitem[Upgren \& Weis(1977)]{upgren1977} Upgren, A.~R. \& Weis, E.~W.\ 1977, \aj, 82, 978. doi:10.1086/112159
\bibitem[van Altena(1966)]{vanaltena1966} van Altena, W.~F.\ 1966, \aj, 71, 482. doi:10.1086/109952
\bibitem[van Altena(1969)]{vanaltena1969} van Altena, W.~F.\ 1969, \aj, 74, 2. doi:10.1086/110768
\bibitem[van Bueren(1952)]{vanbueren1952} van Bueren, H.~G.\ 1952, \bain, 11, 385
\bibitem[van Maanen(1942)]{vanmaanen1942} van Maanen, A.\ 1942, \pasp, 54, 109. doi:10.1086/125413
\bibitem[van Rhijn \& Raimond(1934)]{vanrhijn1934} van Rhijn, P.~J. \& Raimond, J.~J.\ 1934, \mnras, 94, 508. doi:10.1093/mnras/94.6.508
\bibitem[Vos et al.(2017)]{vos2017} Vos, J.~M., Allers, K.~N., \& Biller, B.~A.\ 2017, \apj, 842, 78. doi:10.3847/1538-4357/aa73cf
\bibitem[Wanderley et al.(2023)]{wanderley2023} Wanderley, F., Cunha, K., Souto, D., et al.\ 2023, \apj, 951, 90. doi:10.3847/1538-4357/acd4bd
\bibitem[Weersman(1904)]{weersman1904} Weersman, H.~A.\ 1904, Publications of the Kapteyn Astronomical Laboratory Groningen, 13, 1
\bibitem[Wilson(1948)]{wilson1948} Wilson, R.~E.\ 1948, \apj, 107, 119. doi:10.1086/145000
\bibitem[Wirtz(1902)]{wirtz1902} Wirtz, C.~W.\ 1902, Astronomische Nachrichten, 160, 17. doi:10.1002/asna.19021600202
\bibitem[Zhang et al.(2021)]{zhang2021} Zhang, Z., Liu, M.~C., Best, W.~M.~J., et al.\ 2021, \apj, 911, 7. doi:10.3847/1538-4357/abe3fa
\bibitem[Zhang et al.(2024)]{zhang2024} Zhang, R., Liu, M.~C., \& Zhang, Z.\ 2024, \apj, 960, 105. doi:10.3847/1538-4357/ad083c
\bibitem[Zuckerman \& Becklin(1987)]{zuckerman1987} Zuckerman, B. \& Becklin, E.~E.\ 1987, \apjl, 319, L99. doi:10.1086/184962
\end{thebibliography}
\end{document}